\documentclass[a4paper,twocolumn,11pt,accepted=2025-04-24]{quantumarticle}
\pdfoutput=1

\usepackage[numbers,sort&compress]{natbib}
\usepackage{amsfonts, amsmath, amssymb, mathrsfs, mathtools, amsthm, bbm, bm}
\usepackage{braket}
\usepackage{float}
\usepackage{afterpage}
\usepackage[version=4]{mhchem}
\usepackage[
    labelfont=bf, 
    font=small,
    justification=justified,
    format=plain
    ]{caption}
\usepackage[font=footnotesize,justification=justified]{subcaption}
\usepackage{comment}
\usepackage{tikz}
\usetikzlibrary{shadows,arrows.meta,positioning,backgrounds,fit,shapes}
\usepackage{quantikz}
\usepackage{booktabs, tabularx, colortbl}
\usepackage[linesnumbered,ruled]{algorithm2e}
\usepackage{quantikz}
\usepackage{blochsphere}

\makeatletter
\def\BState{\State\hskip-\ALG@thistlm}
\makeatother

\DeclareMathOperator{\Tr}{Tr}

\newcommand{\fa}{\;\forall\,}
\newcolumntype{Y}{>{\raggedleft\arraybackslash}X}

\theoremstyle{definition}

\newcommand*\circled[1]{\tikz[baseline=(char.base)]{
            \node[shape=circle,draw,inner sep=2pt] (char) {#1};}}

\makeatletter
\renewcommand*\env@matrix[1][*\c@MaxMatrixCols c]{%
  \hskip -\arraycolsep
  \let\@ifnextchar\new@ifnextchar
  \array{#1}}
\makeatother

\definecolor{plasma0}{rgb}{0.299855, 0.009561, 0.631624}
\definecolor{plasma1}{rgb}{0.494877, 0.01199, 0.657865}
\definecolor{plasma2}{rgb}{0.665129, 0.138566, 0.585582}
\definecolor{plasma3}{rgb}{0.798216, 0.280197, 0.469538}
\definecolor{plasma4}{rgb}{0.901807, 0.425087, 0.359688}
\definecolor{plasma5}{rgb}{0.973416, 0.585761, 0.25154}
\definecolor{plasma6}{rgb}{0.993033, 0.77172, 0.154808}

\tikzset{
    my hex/.style={regular polygon, regular polygon sides=6, draw, inner sep=0pt, outer sep=0pt, minimum size=1.15cm,rotate=0},
    small circ/.style={draw, circle, fill=black, inner sep=0pt, minimum size=0.4mm},
    big circ/.style={draw, circle, fill=cyan, inner sep=0pt, minimum size=1.5mm},
    big circ 1/.style={draw, circle, fill=plasma0, inner sep=0pt, minimum size=1.5mm},
    big circ 2/.style={draw, circle, fill=plasma1, inner sep=0pt, minimum size=1.5mm},
    big circ 3/.style={draw, circle, fill=plasma2, inner sep=0pt, minimum size=1.5mm},
    big circ 4/.style={draw, circle, fill=plasma3, inner sep=0pt, minimum size=1.5mm},
    big circ 5/.style={draw, circle, fill=plasma4, inner sep=0pt, minimum size=1.5mm},
    big circ 6/.style={draw, circle, fill=plasma5, inner sep=0pt, minimum size=1.5mm},
    big circ 7/.style={draw, circle, fill=plasma6, inner sep=0pt, minimum size=1.5mm},
    big circ/.style={draw, circle, fill=white, inner sep=0pt, minimum size=1.5mm},
    my circ/.style={draw, circle, fill=black, inner sep=0pt, minimum size=0.9mm},
    my hex select/.style={regular polygon, regular polygon sides=6, inner sep=0pt, outer sep=0pt, minimum size=1.15cm, fill=plasma6!25},
    ancilla select/.style={shape=circle,    minimum width=.3cm, minimum height=.3cm, fill=plasma4!60},
}

\newcommand{\hl}[1]{#1} % uncomment to remove new text highlights

\usepackage[bookmarks=false]{hyperref}

\begin{document}

\title{Accurately Simulating the Time Evolution of an Ising Model with Echo Verified Clifford Data Regression on a Superconducting Quantum Computer}

\author{Tim Weaving}
\affiliation{Centre for Computational Science, Department of Chemistry, University College London, WC1H 0AJ, United Kingdom}
\email{timothy.weaving.20@ucl.ac.uk}
\author{Alexis Ralli}
\affiliation{Centre for Computational Science, Department of Chemistry, University College London, WC1H 0AJ, United Kingdom}
\affiliation{Department of Physics and Astronomy, Tufts University, Medford, MA 02155, USA}
%\email{alexis.ralli.18@ucl.ac.uk}
\author{Peter J. Love}
\affiliation{Department of Physics and Astronomy, Tufts University, Medford, MA 02155, USA}
\affiliation{Computational Science Initiative, Brookhaven National Laboratory, Upton, NY 11973, USA}
%\email{peter.love@tufts.edu}
\author{Sauro Succi}
\affiliation{Center for Life Nano-Neuro Science @ La Sapienza, Italian Institute of Technology, 00161 Roma, Italy}
\affiliation{Department of Mechanical Engineering, University College London, WC1E 7JE, United Kingdom}
\affiliation{Department of Physics, Harvard University, Cambridge, MA 02138, USA}
\author{Peter V. Coveney}
\affiliation{Centre for Computational Science, Department of Chemistry, University College London, WC1H 0AJ, United Kingdom}
\affiliation{Advanced Research Computing Centre, University College London, WC1H 0AJ, United Kingdom}
\affiliation{Informatics Institute, University of Amsterdam, Amsterdam, 1098 XH, Netherlands}
%\email{p.v.coveney@ucl.ac.uk}

% \date{\today}

\begin{abstract}
We present an error mitigation strategy composed of Echo Verification (EV) and Clifford Data Regression (CDR), the combination of which allows one to learn the effect of the quantum noise channel to extract error mitigated estimates for the expectation value of Pauli observables. We analyse the behaviour of the method under the depolarizing channel and derive an estimator for the depolarization rate in terms of the ancilla purity and postselection probability. We also highlight the sensitivity of this probability to noise, a potential bottleneck for the technique. We subsequently consider a more general noise channel consisting of arbitrary Pauli errors, which reveals a linear relationship between the error rates and the estimation of expectation values, suggesting the learnability of noise in EV by regression techniques. Finally, we present a practical demonstration of Echo Verified Clifford Data Regression (EVCDR) on a superconducting quantum computer and observe accurate results for the time evolution of an Ising model over spin-lattices consisting of up to 35 sites and \hl{circuit depths up to 173 entangling layers.}
\end{abstract}

\maketitle

\section{Introduction}\label{sec:intro}

The greatest challenge facing noisy intermediate-scale quantum (NISQ) computation is the mitigation of error, which manifests as imperfection in quantum logical operations or during readout. Measurement-Error Mitigation \cite{bravyi2021mitigating, nation2021scalable}, whereby one rebalances the measurement distribution given probabilities that a particular qubit erroneously flips $\ket{0} \leftrightarrow \ket{1}$, addresses readout errors. Gate error is harder to mitigate and may be categorized either as \textit{coherent}, arising for example as imprecision in gate rotation angles and device miscalibration, or \textit{incoherent}, representing a coupling of the system with its environment. Coherent error is hard to suppress since it leads to a systemic bias in the output, while incoherent error is stochastic in nature and its average effect is often well-described by the \textit{depolarizing channel} \cite{urbanek2021mitigating}; this is particularly appropriate with increasing system dimension $d$ and circuit complexity. 

Considerable effort has gone into developing means of converting gate errors from coherent to incoherent, most notably through practices such as \textit{Pauli Twirling}~\cite{geller2013twirling, cai2019twirling} and \textit{Randomized Compiling}~\cite{wallman2016noise, hashim2020randomized}. This directs quantum noise into the Pauli channel that is subsequently mitigated by techniques such as \textit{Zero Noise Extrapolation} (ZNE)~\cite{li2017efficient, temme2017error, endo2018practical, kandala2019error, giurgica2020digital, he2020zero, mari2021extending, maupin2023error, weaving2025contextual} and \textit{Echo Verification} (EV)~\cite{o2021error, cai2021resource, weaving2023benchmarking, kiss2024quantum, schiffer2024virtual}, the latter of which is the focus for this work. EV, also referred to as \textit{Dual-State Purification} \cite{huo2022dual}, may be viewed as a NISQ-friendly variant of \textit{Virtual Distillation} (VD)~\cite{huggins2021virtual, liu2024virtual}, whereby one induces the product $\rho^M$ over $M$ copies of a quantum state $\rho$, the effect being that the dominant pure component will be amplified in the spectral decomposition, while lower lying contributions will be exponentially suppressed. EV is related to the $M=2$ case of VD, although it is not equivalent owing to its dual structure.

\begin{figure}[b!]
    \centering
    \resizebox{\linewidth}{!}{\input{dsp_circ}}
    \caption{Echo Verification circuit for estimating the expectation value $\braket{V} = \bra{\varphi} V \ket{\varphi}$ where $\ket{\varphi}=U \ket{0}$ and $U, V$ are unitary operators; for example, $V$ might be a Pauli operator and $U$ some ansatz circuit. This is related to a Hadamard test for the measurement of $U^\dag V U$ with respect to the zero state, but is distinguished through the addition of a postselection procedure. This projection onto the zero-space drives the ancilla qubit into the state \eqref{depolarized_rho} from which one may extract a noise-mitigated estimate of $\braket{V}$.}
    \label{fig:dsp_circ}
\end{figure}

\begin{figure}[b!]
    \centering
    \resizebox{0.9\linewidth}{!}{\input{light_cone_circ}}
    \caption{Pictorial demonstration of how the Echo Verification structure naturally identifies the light-cone for an observable. Gates lying outside the observable support (indicated here with shading) are cancelled under application of the inverse circuit and thus can lead to significant reductions in gate counts for highly localized observables. For the purposes of near-Clifford noise learning we permit a handful of rotation gates within the light-cone to assume non-Clifford rotation angles.}
    \label{fig:light-cone}
\end{figure}

In EV, one prepares the pure state
\begin{equation}\label{standard_ev_psi}
    \ket{\psi} = \frac{1}{\sqrt{2}}\Big(\ket{\bm{0}}^{\mathrm{sys}}\otimes\ket{+}^{\mathrm{anc}}+U^\dag V U \ket{\bm{0}}^{\mathrm{sys}}\otimes\ket{-}^{\mathrm{anc}}\Big)
\end{equation}
before postselecting on zero-measurements in the system register. Here $\ket{\pm}^{\mathrm{anc}}$ are basis states of a single ancilla qubit. This induces the desired expectation value $\braket{V} = \bra{\bm{0}} U^\dag V U \ket{\bm{0}}^{\mathrm{sys}}$ on the ancilla qubit, whose noise-free reduced state is  $\rho^{\mathrm{anc}} = \ket{\psi}\bra{\psi}^{\mathrm{anc}}$ for
\begin{equation}\label{noiseless_ancilla}
    \ket{\psi}^{\mathrm{anc}} = \frac{1}{\sqrt{2p_{\bm{0}}}} \big(\ket{+}^{\mathrm{anc}} + \braket{V}\ket{-}^{\mathrm{anc}}\big)
\end{equation}
where $p_{\bm{0}} = |(\bra{\bm{0}} \otimes I ) \ket{\psi}|^2 = \frac{1+\braket{V}^2}{2}$ is the probability of postselecting the zero state in the system register, depicted in Figure \ref{fig:dsp_circ}. Note that $p_{\bm{0}}\geq 1/2$. An advantage of EV over alternative error mitigation techniques is that it naturally identifies the light-cone \cite{tran2019locality, leone2022practical} of the measured observable due to a cancellation of unsupported gates owing to its dual structure, illustrated in Figure \ref{fig:light-cone}. This avoids the need to use techniques such a light-cone tracing \cite{tran2023locality} or the out-of-time-order correlator (OTOC) \cite{mi2021information}.

The quantity $\braket{V}$ may then be extracted from the ancilla qubit by performing $X$- and $Z$-basis measurements, which yield estimators $\mathcal{E}_X, \mathcal{E}_Z$ satisfying
\begin{equation}
\begin{aligned}
    \mathbb{E}(\mathcal{E}_X) ={} & \Tr{(X\rho^{\mathrm{anc}})} = \frac{1-\braket{V}^2}{1+\braket{V}^2} \\ 
    \mathbb{E}(\mathcal{E}_Z) ={} & \Tr{(Z\rho^{\mathrm{anc}})} = \frac{2\braket{V}}{1+\braket{V}^2}
\end{aligned}
\end{equation}
and thus an unbiased estimator
\begin{equation}\label{EV_standard_estimator}
    \mathcal{E}_{\mathrm{EV}} = \mathcal{E}_Z(1+\mathcal{E}_X)^{-1}
\end{equation}
of the expectation value $\braket{V} = \mathbb{E}(\mathcal{E}_{\mathrm{EV}})$. By observing that $\mathcal{E}_X^2+\mathcal{E}_Z^2=1$, one may rewrite the standard EV estimator \eqref{EV_standard_estimator} in multiple ways \cite{weaving2023benchmarking}:
\begin{equation}\label{EV_variant}
\begin{aligned}
    \mathcal{E}^{Z\mathrm{-bias}}_{\mathrm{EV}} ={} & \mathcal{E}_Z \big(1+\sqrt{1-\mathcal{E}_Z^2} \big)^{-1} \\
    \mathcal{E}^{X\mathrm{-bias}}_{\mathrm{EV}} ={} & \mathrm{sign}(\mathcal{E}_Z) \sqrt{(1-\mathcal{E}_X)(1+\mathcal{E}_X)^{-1}}.
\end{aligned}
\end{equation}

\begin{figure*}
    \centering
    \resizebox{\linewidth}{!}{
        \input{ev-cdr_flowchart}
    }
    \caption{Overview of the Echo Verified Clifford Data Regression (EVCDR) framework. The results obtained from a standard Echo Verification routine are biased with a linear noise damping factor that is learned via the Clifford Data Regression procedure. This is achieved by randomly sampling near-Clifford training circuits by approximating all but a few of the rotation gates by their nearest Clifford counterpart (for example, by rounding gate angles to an integer multiple of $\frac{\pi}{2}$). By evaluating the training circuits both classically and on the quantum hardware we may learn the effect of the underlying noise channel and subsequently suppress error in the final energy estimate. We may then reflect estimates from the true non-Clifford circuit through the Clifford fitting curves to obtain CDR mitigated estimates that are then combined in the same way as the standard EV estimator.}
    \label{fig:EV_flowchart}
\end{figure*}

The estimator $\mathcal{E}^{Z\mathrm{-bias}}_{\mathrm{EV}}$ requires only $Z$-basis measurements to obtain $\braket{V}$ when $\mathcal{E}_x\geq 0$, while $\mathcal{E}^{X\mathrm{-bias}}_{\mathrm{EV}}$ cannot be determined with $X$-basis measurements alone, since the sign information is not accessible; supplementary $Z$-basis readings are necessary to decide the $\pm1$ coefficient, although significantly fewer shots suffice to indicate this sign. These alternative forms are not the same in practice due to differences in how the noise channel affects $X/Z$ measurements on the ancilla, which is analysed formally in Section \ref{pauli_channel}. As such, we may find significant disparity between $\mathcal{E}_{\mathrm{EV}}, \mathcal{E}^{X\mathrm{-bias}}_{\mathrm{EV}}, \mathcal{E}^{Z\mathrm{-bias}}_{\mathrm{EV}}$, with one possibly displaying better performance than the others.

In this work we study the EV protocol under application of several quantum channels so that we can understand how noise propagates through to the ancilla qubit and subsequently identify improvements to the technique. In Section \ref{sec:EV_depolarizing} we consider the simplest setting of depolarizing noise; while it was previously argued that EV guarantees $\geq50$\% retention of circuit samples through the postselection procedure \cite{huo2022dual}, we highlight a linear decay in the success probability $p_{\bm{0}}$ with the rate of depolarization to a minimum of $2/d$. This explains observations of $p_{\bm{0}}<0.5$ in our previous experimental benchmarking on IBM superconducting hardware \cite{weaving2023benchmarking}. We then derive an estimator for the depolarization rate, a quantity that is ordinarily inaccessible, in terms of the postselection probability and ancilla purity, albeit with an adverse sampling overhead. 

In Section \ref{pauli_channel} we study a more general noise channel consisting of arbitrary Pauli errors and discover a linear relationship between the error rates and noisy expectation values. This reveals the learnability of the effect of Pauli error in EV, which one may probe using \textit{Clifford Data Regression} (CDR)~\cite{czarnik2021error} to further suppress bias in the estimator through near-Clifford training circuits, as investigated in Section \ref{clifford_learning}. We also note in Section \ref{sec:noise_model_limits} the limitations of the noise model used in our analyses, given that application of the noise channel is delayed to the end of the circuit as in Figure \ref{fig:dsp_circ}.

The addition of CDR to the EV method produces a combined error mitigation strategy that is greater than the sum of its parts, which we shall henceforth refer to as \textit{Echo~Verified~Clifford~Data~Regression}~(EVCDR). Finally, in Section \ref{sec:results} we present a practical showcase of the EVCDR method, in which we accurately simulate the time evolution of Ising models on heavy-hex spin-lattices consisting of 21 and 35 qubits.

\section{Echo Verification under Depolarizing Noise}\label{sec:EV_depolarizing}

The standard formulation of Echo Verification introduced in Section \ref{sec:intro} does not take into account the effects of systematic noise; our goal is to modify the estimator \eqref{EV_standard_estimator} to remove any bias arising from the presence of depolarizing noise. This is an idealized model of quantum noise and previous works have attempted to learn the depolarization rate in order to mitigate error, for example with Clifford estimation circuits \cite{urbanek2021mitigating}.

We model this by passing the pure state $\rho = \ket{\psi}\bra{\psi}$ as per Equation \ref{standard_ev_psi} through the quantum channel 
\begin{equation}\label{depolarizing_channel}
    \Phi_\delta(\rho)=(1-\delta)\rho + \frac{\delta}{d}I^{\otimes N},
\end{equation} 
before application of the postselection projector $M_0 = \ket{\bm{0}}\bra{\bm{0}}_{\mathrm{sys}}\otimes I_{\mathrm{anc}}$ with probability $p_{\bm{0}}(\delta) = \Tr{(M_0 \Phi_\delta(\rho))}$ and observe the propagation of noise to the ancilla qubit. Following this process, the mixed state of the ancilla qubit is
\begin{equation}\label{depolarized_rho}
\begin{aligned}
    \rho^{\mathrm{anc}}_{\delta} \coloneqq & \frac{\Tr_{\mathrm{sys}}{(M_0 \Phi_\delta(\rho))}}{\Tr{(M_0 \Phi_\delta(\rho))}} \\
    ={} & \frac{1-\delta}{2p_{\bm{0}}(\delta)} \Big[ \ket{+}\bra{+} + \braket{V}^2 \ket{-}\bra{-} \\ & \hspace{4mm} + \braket{V} \big( \ket{+}\bra{-} + \ket{-}\bra{+} \big) \Big] + \frac{\delta}{d p_{\bm{0}}(\delta)} I.
\end{aligned}
\end{equation}
By asserting $\Tr{(\rho^{\mathrm{anc}}_{\delta})} = 1$ we find there is a linear dependency between the depolarization strength and the zero postselection probability
\begin{equation}\label{p0_depolarizing}
    p_{\bm{0}}(\delta) = p_{\bm{0}} \cdot (1-\delta)+\frac{2\delta}{d}
\end{equation}
where
\begin{equation}
    p_{\bm{0}} = p_{\bm{0}}(0) = \frac{1+\braket{V}^2}{2} \geq \frac{1}{2}    
\end{equation}
is the probability of successfully postselecting the zero state in the system register for the noiseless scenario \eqref{noiseless_ancilla}; this has also been noted in another work \cite{kiss2024quantum}. Therefore, it is possible to reduce the probability of success to any value $p_{\bm{0}}(\delta) = p_{\bm{0}}^\prime$ with $p_{\bm{0}} > p_{\bm{0}}^\prime > \frac{2}{d}$ given a sufficiently high rate of depolarization $\delta>\frac{p_{\bm{0}} - p_{\bm{0}}^\prime}{p_{\bm{0}}-2/d}$. In the worst case scenario of $\braket{V}=0$ the probability of success will always be lower than $50\%$ for non-zero depolarization, whereas for $|\braket{V}|=1$ we may tolerate $\delta \leq \frac{1}{2}(\frac{1}{1-2/d}) \approx 0.5$ for large dimensions $d$ and still retain $\geq50\%$ of the samples.

\begin{figure}[b!]
    \centering
    \includegraphics[width=\linewidth]{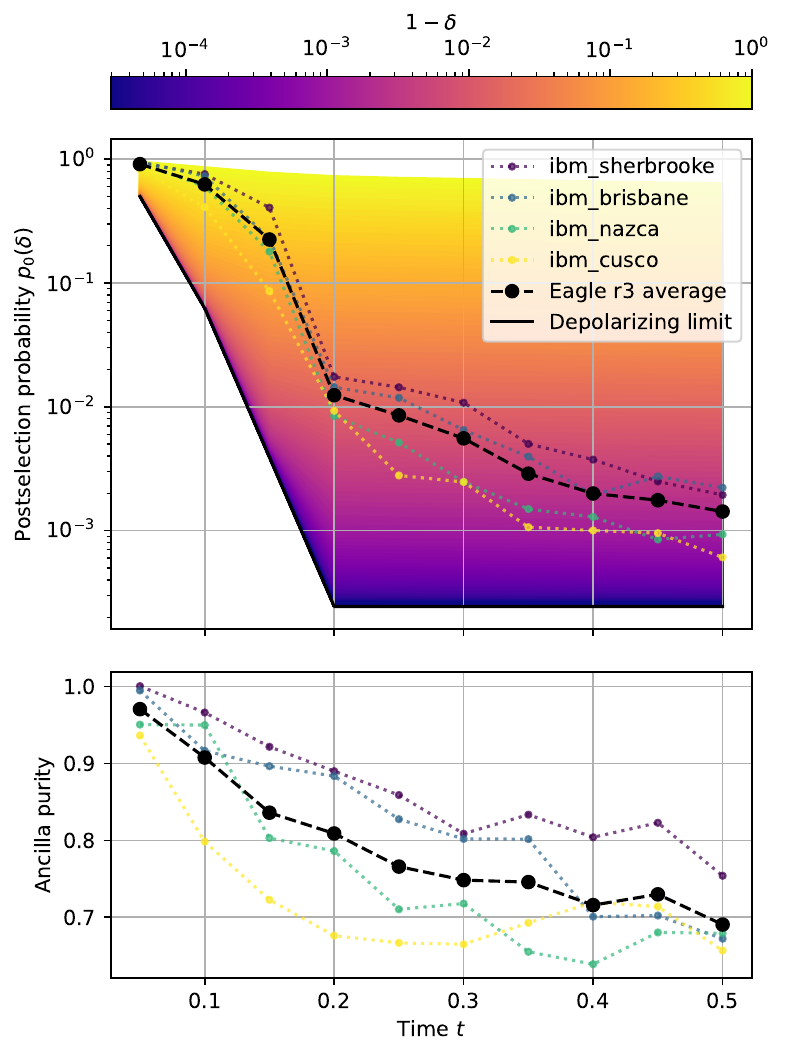}
    \caption{Postselection probability \eqref{p0_depolarizing} and ancilla \eqref{depolarized_rho} purity over time evolution of a 12-qubit Ising ring, averaged across the Eagle r3 QPUs \textit{ibm\_cusco}, \textit{ibm\_nazca}, \textit{ibm\_brisbane} and \textit{ibm\_sherbrooke}; six circuit instances were tiled across these 127-qubit chips. The circuit depth increases linearly with the time step, thus the depolarization rate also increases and therefore we observe a decay in $p_{\bm{0}}(\delta)$ which will converge on the depolarizing limit of $\frac{2}{d}$. This limit is not constant as the light-cone is small for early times and is saturated from $t=0.2$ onwards.}
    \label{fig:p0}
\end{figure}

In Figure \ref{fig:p0} we observe the decay in postselection probability \eqref{p0_depolarizing} for a selection of Eagle r3 chips as the circuit depth increases under time evolution of a 12-qubit Ising ring. \hl{Here, depth is taken to mean the number of echoed cross-resonance (ECR) layers, the native entangling gate on the hardware.} This also serves to indicate the quality of each chip, with \textit{ibm\_sherbrooke} yielding the greatest $p_{\bm{0}}(\delta)$ and $\gamma(\rho^{\mathrm{anc}})$ values, whereas \textit{ibm\_cusco} is the poorest performer here; this is in agreement with IBM's error per layered gate (EPLG) metric, with stated values of $1.7\%$ and $5.9\%$, respectively at the time of writing. The convergence of $p_{\bm{0}}(\delta)$  on the depolarizing limit of $\frac{2}{d}$ indicates that the noise channel is itself converging on the depolarizing channel, demonstrating that it describes the average effect of noise appropriately. This is problematic as it seems as though we will quickly saturate the depolarizing limit for larger systems and deep circuits, meaning the sampling overhead needs to scale inversely with the postselection probability $p_{\bm{0}}(\delta)$ \eqref{p0_depolarizing}. More precisely, for an error tolerance $\epsilon > 0$ one needs to retain $\mathcal{O}(\epsilon^{-2})$ samples through the postselection procedure and so to achieve $p_{\bm{0}}(\delta) \cdot n_{\mathrm{sample}} = \mathcal{O}(\epsilon^{-2})$ requires 
\begin{equation}
n_{\mathrm{sample}}~=~\mathcal{O}\big( \epsilon^{-2} (1-\delta)^{-1} \big).
\end{equation}
In Appendix \ref{sec:volumetric} we observe the postselection probability obtained with a fixed shot budget $n_{\mathrm{sample}}=8,192$ for Loschmidt echos $UU^\dag$ where $U$ is averaged over unitaries selected uniformly at random. We note the structure $UU^\dag$ is closely related to EV, being the trivial case where $V=I$ in Figure \ref{fig:dsp_circ}.

We now wish to derive a relation between the probability of success $p_{\bm{0}}(\delta)$ and the ancilla purity $\gamma(\rho^{\mathrm{anc}}_{\delta})$, with the goal of estimating the rate of depolarization $\delta$. In order to do so, we may write the depolarized form \eqref{depolarized_rho} of the ancillary system in terms of the noiseless expression \eqref{noiseless_ancilla}; recalling $\rho^{\mathrm{anc}} = \ket{\psi}\bra{\psi}^{\mathrm{anc}}$ where $\ket{\psi}^{\mathrm{anc}} = \frac{1}{\sqrt{2p_{\bm{0}}}} \big(\ket{+}^{\mathrm{anc}} + \braket{V}\ket{-}^{\mathrm{anc}}\big)$ and by introducing the quantity $\alpha(\delta) \coloneqq \frac{\delta}{dp_{\bm{0}}(\delta)}$, one finds
\begin{equation}
\begin{aligned}
    \rho^{\mathrm{anc}}_{\delta} 
    ={} & \big(1-2\alpha(\delta)\big) \rho^{\mathrm{anc}} + \alpha(\delta) I.
\end{aligned}
\end{equation}

Using the fact that $\gamma(\rho^{\mathrm{anc}})=1$, it can be seen that the purity of the ancilla state under depolarizing noise is quadratic in the newly introduced parameter $\alpha(\delta)$:
\begin{equation}
    \gamma(\rho^{\mathrm{anc}}_{\delta}) = 1-2\alpha(\delta)+2\alpha(\delta)^2.
\end{equation}
Solving this yields an expression for the depolarization rate in terms of the ancilla purity and postselection probability:
\begin{equation}
    \delta = \frac{dp_{\bm{0}}(\delta)(1-\gamma(\rho^{\mathrm{anc}}_{\delta}))}{1+\sqrt{2\gamma(\rho^{\mathrm{anc}}_{\delta})-1}}.
\end{equation}
The significance of this result is that $\delta$ is not ordinarily accessible. However, we have shown that it is possible to approximate the depolarization rate from a number of quantities we are able to estimate through circuit sampling; the probability of success $p_{\bm{0}}(\delta)$ is simply the proportion of measurements that survive the postselection procedure and the ancilla purity $\gamma(\rho^{\mathrm{anc}}_{\delta})$ may be calculated by performing state tomography on the ancilla qubit, requiring just $X,Y,Z$ measurements. 

\begin{figure}[b!]
    \centering
    \includegraphics[width=\linewidth]{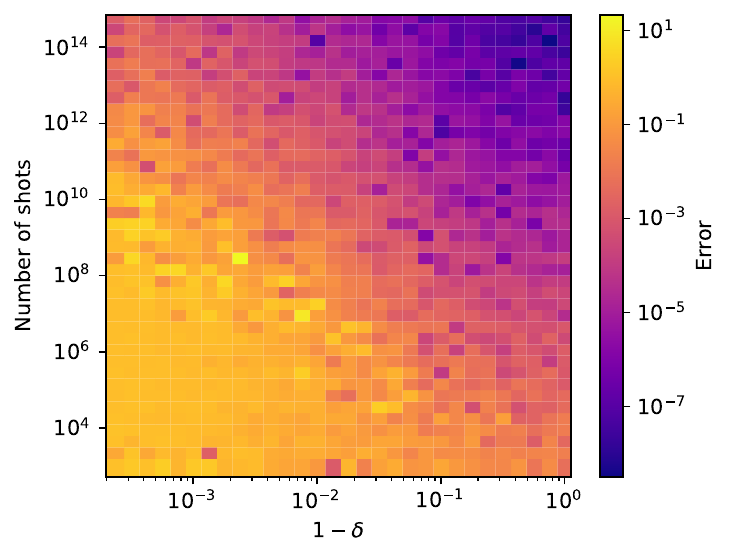}
    \caption{$\mathcal{E}_{\mathrm{EV}}^\delta$ \eqref{EV_tolerant_estimator} error as a function of sampling budget and depolarization strength $\delta$ in an idealized model of depolarizing noise for an Ising ring. It may be seen in the top left of this plot that, for high levels of depolarizing noise with $\delta \approx 1$, a very large number of circuit shots is required to extract the expectation value faithfully.}
    \label{fig:error_by_shots_and_delta}
\end{figure}

It is then possible to counteract the bias introduced by depolarizing noise, since
\begin{equation}
\begin{aligned}
    \Tr{(X\rho^{\mathrm{anc}}_{\delta})} ={} & \frac{1-\delta}{p_{\bm{0}}(\delta)} \cdot \frac{1 - \braket{V}^2}{2}  \\
    \Tr{(Z\rho^{\mathrm{anc}}_{\delta})} ={} & \frac{1-\delta}{p_{\bm{0}}(\delta)} \cdot \braket{V}
\end{aligned}
\end{equation}
and therefore the standard EV estimator \eqref{EV_standard_estimator} yields
\begin{equation}
    \mathbb{E}(\mathcal{E}_{\mathrm{EV}}) = \frac{\Tr{(Z\rho^{\mathrm{anc}}_{\delta})}}{1+\Tr{(Z\rho^{\mathrm{anc}}_{\delta})}} = \braket{V}\cdot\frac{1-\delta}{1-\delta(1-2/d)}.
\end{equation}
Finally, we obtain a depolarization-tolerant estimator
\begin{equation}\label{EV_tolerant_estimator}
    \mathcal{E}_{\mathrm{EV}}^{\delta} = \mathcal{E}_{\mathrm{EV}}\bigg(1 + \frac{2\delta}{d(1-\delta)}\bigg)
\end{equation}
such that $\mathbb{E}(\mathcal{E}_{\mathrm{EV}}^\delta) = \braket{V}$ regardless of the depolarization rate $\delta$. 

The difficulty is estimating $\delta$ with sufficiently high precision, given that the dimension $d$ grows exponentially and thus dominates this expression. The consequence is the sampling overhead scales exponentially, which is demonstrated in Figure \ref{fig:error_by_shots_and_delta} where we observe the estimator \eqref{EV_tolerant_estimator} error as a function of sampling budget and error rate for a 12-qubit Ising ring in an idealized model of depolarizing noise. This fact renders the above approach to EV infeasible in practice. 

The next section presents analysis of EV through an arbitrary Pauli channel to assess whether some other feature of the method might be exploited to derive greater suppression of estimator bias than the standard approach, while maintaining similar levels of practicality.

\section{Echo Verification Under Pauli Noise}\label{pauli_channel}

We now consider the setting of arbitrary Pauli noise, defined by a channel $\Phi: \rho \mapsto \sum_{i=1}^{d^2} \lambda_i P_i \rho P_i$ with error probabilities $\lambda_i$ such that $\sum_i \lambda_i = 1$. We explicitly partition the qubits into the system $\mathscr{H}^{(\mathrm{sys})}$ and ancilla $\mathscr{H}^{(\mathrm{anc})}$ registers so that $\mathscr{H} = \mathscr{H}^{(\mathrm{sys})} \otimes \mathscr{H}^{(\mathrm{anc})}$ and we may write a Pauli $P \in \mathscr{B}(\mathscr{H})$ across this division $P = P^{(\mathrm{sys})} \otimes P^{(\mathrm{anc})}$. The ancillary system contains just a single qubit and therefore $P^{(\mathrm{anc})} \in \{I,X,Y,Z\}$; in Appendix \ref{sec:multi_ancilla_EV} we discuss two approaches to multi-ancilla Echo Verification and its feasibility.

Under this channel, as derived in Appendix \ref{sec:arb_pauli_noise}, the mixed state of the ancilla qubit in Echo Verification is
\begin{equation}\label{eq:anc_rho}
\begin{aligned}
    \rho^{\mathrm{anc}} 
    \coloneqq{} & \frac{\Tr_{\mathrm{sys}}(M_0 \Phi(\rho))}{\Tr(M_0 \Phi(\rho))} \\
    ={} & \sum_{i=1}^{\frac{d^2}{4}} \big\{ \lambda_i^{I}\rho_i + \lambda_i^{X} X\rho_iX + \lambda_i^{Y} Y\rho_iY + \lambda_i^{Z} Z\rho_iZ \big\}
\end{aligned}
\end{equation}
where
\begin{equation}
\begin{aligned}
    \rho_i ={} & \frac{1}{2p_{0|i}} \Big[ \delta_i^{Z} \ket{+}\bra{+} + |\Gamma_i|^2 \ket{-}\bra{-} \\ & \hspace{14mm} + \delta_i^{Z} \braket{V} \big(\ket{+}\bra{-} + \ket{-}\bra{+}\big)\Big]
\end{aligned}
\end{equation}
for $\Gamma_i = \bra{0} U^\dag V U P_i^{(\mathrm{sys})} \ket{0}$ and $\braket{V} = \bra{0}U^\dag V U \ket{0}$. We have that $\delta_i^{Z}$ is one when $P_i^{(\mathrm{sys})}$ is of $Z$-type error only, while it is zero if the system error $P_i^{(\mathrm{sys})}$ contains $X$ or $Y$ operations. This suppression arises from the postselection procedure where such off-diagonal error vanishes. Note that, while it looks like this statement asserts a product form for the noise channel between system and ancilla, this is not the case since $\rho_i$ changes for each summand in Equation \eqref{eq:anc_rho}, and the noise rates $\lambda_i^{Q}$ for $Q \in \{I,X,Y,Z\}$ correspond with the same Pauli error rates $\lambda_i$ from before, but partitioned such that $P_i^{(\mathrm{anc})} = Q$, i.e. the error rate associated with the Pauli error $P_i = P_i^{(\mathrm{sys})} \otimes Q$ that applies $Q$ on the ancilla. Refer to Equations \eqref{projected_phi_rho} and \eqref{collect_anc_rates} of Appendix \ref{sec:arb_pauli_noise} for further details.

By asserting $\Tr{(\rho^{\mathrm{anc}})}=1$ we may now infer $p_{\bm{0}|i} = \frac{\delta_i^Z+|\Gamma_i|^2}{2}$, which represents the conditional probability of postselecting zero given that the Pauli error $P_i^{(\mathrm{sys})}$ has occurred; overall we have $p_{\bm{0}} = \frac{1}{2}\sum_{i=1}^{d^2/4} (\lambda^{I}_i+\lambda^{X}_i+\lambda^{Y}_i+\lambda^{Z}_i)(\delta_i^{Z}+|\Gamma_i|^2)$.

Observing $\Tr(PQ\rho_iQ) = (-1)^{\delta_{0,\{P,Q\}}} \Tr(P\rho_i)$, measurement of the ancilla qubit in the $Z$-basis under this more general noise setting yields
\begin{equation}
\begin{aligned}
    \Tr{(Z\rho^{\mathrm{anc}})} ={} & \sum_{i=1}^{\frac{d^2}{4}} \Tr{(Z\rho_i)} (\lambda_i^{I}-\lambda_i^{X}-\lambda_i^{Y}+\lambda_i^{Z}) \\
    ={} & \sum_{i=1}^{\frac{d^2}{4}} \frac{2\braket{V}\delta_i^Z}{\delta_i^Z+|\Gamma_i|^2} (\lambda_i^{I}-\lambda_i^{X}-\lambda_i^{Y}+\lambda_i^{Z}) \\
    ={} & \frac{2\braket{V}}{1+\braket{V}^2} \cdot \Lambda_Z
\end{aligned}
\end{equation}
where
\begin{equation}\label{lambdaZ}
    \Lambda_Z = \sum_{\substack{i=1\\P_i^{(\mathrm{sys})} \in \{I,Z\}^{\otimes N}}}^{\frac{d^2}{4}} (\lambda_i^{I}-\lambda_i^{X}-\lambda_i^{Y}+\lambda_i^{Z}).
\end{equation}
The final line follows as $\Gamma_i = \braket{V}$ whenever $P_i^{(\mathrm{sys})}$ is strictly of $Z$-type error, which the Kronecker factor $\delta_i^{Z}$ isolates. As we can see, the $Z$-basis measurement yields the same as the noiseless scenario up to some damping factor $\Lambda_Z$, which comprises a sum over rates of Pauli errors that are diagonal in the system register.

Considering now the $X$-basis, we have
\begin{equation}
\begin{aligned}
    \Tr{(X\rho^{\mathrm{anc}})} ={} & \sum_{i=1}^{\frac{d^2}{4}} \Tr{(X\rho_i)} (\lambda_i^{I}+\lambda_i^{X}-\lambda_i^{Y}-\lambda_i^{Z}) \\
    ={} & \sum_{i=1}^{\frac{d^2}{4}} \frac{\delta_i^Z-|\Gamma_i|^2}{\delta_i^Z+|\Gamma_i|^2} (\lambda_i^{I}+\lambda_i^{X}-\lambda_i^{Y}-\lambda_i^{Z}) \\
    ={} & \frac{1-\braket{V}^2}{1+\braket{V}^2} \cdot \Lambda_X - \Omega_X
\end{aligned}
\end{equation}
where
\begin{equation}\label{lambdaX}
\begin{aligned}
    \Lambda_X ={} & \sum_{\substack{i=1\\P_i^{(\mathrm{sys})} \in \{I,Z\}^{\otimes N}}}^{\frac{d^2}{4}} (\lambda_i^{I}+\lambda_i^{X}-\lambda_i^{Y}-\lambda_i^{Z}), \\
    \Omega_X ={} & \sum_{\substack{i=1\\P_i^{(\mathrm{sys})} \not\in \{I,Z\}^{\otimes N}}}^{\frac{d^2}{4}} (\lambda_i^{I}+\lambda_i^{X}-\lambda_i^{Y}-\lambda_i^{Z}).
\end{aligned}
\end{equation}
Once again, the Pauli channel affects the expectation value linearly, however in addition to the damping factor $\Lambda^X$ we find an additive error term $\Omega_X$ since $(\delta_i^Z-|\Gamma_i|^2)/(\delta_i^Z+|\Gamma_i|^2) = -1$ when $P_i^{(\mathrm{sys})}~\not\in~\{I,Z\}^{\otimes N}$ (where $\delta_i^Z=0$) and therefore these off-diagonal error contributions do not vanish as they did in the $Z$-basis measurement. For this reason, one could argue that the $Z$-biased variant $\mathcal{E}^{Z\mathrm{-bias}}_{\mathrm{EV}}$ \eqref{EV_variant} of the EV estimator should perform optimally since all the off-diagonal Pauli error is theoretically suppressed, while the standard $\mathcal{E}_{\mathrm{EV}}$ and $X$-biased $\mathcal{E}^{X\mathrm{-bias}}_{\mathrm{EV}}$ formulations may retain some off-diagonal noise contributions that propagate through to the ancilla qubit and only contaminate the $X$-basis measurements.

Finally, in the $Y$-basis
\begin{equation}
    \Tr{(Y\rho^{\mathrm{anc}})} = \sum_{i=1}^{\frac{d^2}{4}} \Tr{(Y\rho_i)} (\lambda_i^{I}-\lambda_i^{X}+\lambda_i^{Y}-\lambda_i^{Z}) = 0,
\end{equation}
and thus Pauli errors alone do not lead to non-zero $Y$ expectation value and must therefore be a consequence of sampling noise or errors outside the Pauli channel. Note that, by fixing $\lambda_i^{I}=\lambda_i^{X}=\lambda_i^{Y}=\lambda_i^{Z}=\frac{\delta}{d^2} \fa i\in\{1,\dots,\frac{d^2}{4}\}$ except for $\lambda_1^{I} = 1-\delta+\frac{\delta}{d^2}$, we obtain $\Lambda_X = \Lambda_Z = 1-\delta$ and $\Omega_X=0$, thus the depolarizing channel is recovered and yields the same result as in Section \ref{sec:EV_depolarizing}.

Also note that the purity of the ancilla qubit satisfies $\gamma(\rho^{\mathrm{anc}})^2 \approx~\Tr{(X\rho^{\mathrm{anc}})}^2 + \Tr{(Y\rho^{\mathrm{anc}})}^2 + \Tr{(Z\rho^{\mathrm{anc}})}^2$ where $\Tr{(Y\rho^{\mathrm{anc}})}=0$ and therefore renormalizing $\mathcal{E}_X \mapsto \mathcal{E}_X/\gamma(\rho^{\mathrm{anc}}), \mathcal{E}_Z \mapsto \mathcal{E}_Z/\gamma(\rho^{\mathrm{anc}})$ recovers the property $\mathcal{E}_X^2 + \mathcal{E}_Z^2 = 1$; this leads to a modified EV estimator that we shall refer to later as \textit{purity normalized}.

The above observation, that application of the Pauli channel in Echo Verification manifests as a linear relationship between the error rates and the estimation of expectation values, leads us to conclude that we may attempt to learn the noise factors $\Lambda_Z,\Lambda_X,\Omega_X$ through Clifford estimation circuits \cite{czarnik2021error, urbanek2021mitigating}; we investigate this in Section \ref{clifford_learning}.

\subsection{Limitation of the Noise Model}\label{sec:noise_model_limits}

We note that the noise model used for out analyses makes the assumption that all noise can be applied to the final density matrix, illustrated in Figure \ref{fig:dsp_circ}, rather than on a gate-by-gate basis. This really only holds true for Pauli noise applied to a Clifford circuit, therefore given an arbitrary non-Clifford circuit this simplification introduces some approximation error to the model. To illustrate this, consider a Pauli noise channel $\Phi: \rho \mapsto \sum_{i=1}^{d^2} \lambda_i P_i \rho P_i$ and a Clifford gate $C$. Then there exists a permutation map $\pi:\mathbb{N}_{d^2} \mapsto \mathbb{N}_{d^2}$ such that $CP_iC^\dag = P_{\pi(i)} \;\forall i\in\mathbb{N}_{d^2}$, and therefore $CP_i = P_{\pi(i)}C$.

Applying the noise channel before application of the gate $C$ yields
\begin{equation}
\begin{aligned}
    C \Phi(\rho) C^\dag ={} & \sum_{i=1}^{d^2} \lambda_i C P_i \rho P_i C^\dag \\
    ={} & \sum_{i=1}^{d^2} \lambda_i P_{\pi(i)} C \rho C^\dag P_{\pi(i)} \\
    ={} & \sum_{i=1}^{d^2} \lambda_{\pi^{-1}(i)} P_{i} C \rho C^\dag P_{i} \\
    ={} & \Phi^\prime(C \rho C^\dag)
\end{aligned}
\end{equation}
where the modified channel $\Phi^\prime$ differs from $\Phi$ by a permutation of the error rates associated with each Pauli error. Note the restricted setting of depolarizing noise, as investigated in Section \ref{sec:EV_depolarizing}, is the one in which we have $\Phi^\prime \equiv \Phi$. We can apply this recursively to delay application of the noise channel to the end of the circuit.

This does not hold in general and so future work may wish to expand the EV analyses presented here to a more general noise model. In the next section we investigate (near) Clifford approximations to the target circuit, motivated by the findings of Section \ref{pauli_channel}; refinement of the noise model may lead to even more effective EV methods.

\section{Learning the Effect of Noise in Echo Verification Through Near-Clifford Estimation Circuits}\label{clifford_learning}

The \textit{Clifford group} consists of operators $C \in \mathscr{B}(\mathscr{H})$ satisfying $CPC^\dag \in \mathscr{P}_N \fa P \in \mathscr{P}_N$; in other words, Clifford operations are those that map the Pauli group back onto itself, meaning it \textit{normalizes} the Pauli group. Note also that the Pauli operators are themselves Clifford, so $\mathscr{P}_N \subset \mathscr{C}(\mathscr{H})$; additional gates that fall within the Clifford group are $S$, $H$ and CNOT, which are actually sufficient to generate the full group. 

Rotations $R_{x/y/z}(\theta)~=~\exp{(-i\frac{\theta}{2} X/Y/Z)}$ are generally non-Clifford, but they become Clifford at integer multiples of $\frac{\pi}{2}$, i.e. with rotation angles $\theta = \frac{k\pi}{2}$ with $k \in \mathbb{Z}$. Note that some works adopt a different rotation convention and therefore angles might differ.

An important consequence of a circuit being comprised of just Clifford gates is that it is efficiently simulable by classical means \cite{gottesman1997stabilizer}; as an example, the \textit{Symmer} Python package \cite{symmer2022} includes a basic Clifford simulator that can evaluate Hamiltonian expectation values over thousands of qubits within seconds. Furthermore, one may permit a small number of non-Clifford gates within the circuit while still maintaining classical simulability; it is this fact that we shall exploit to learn the Pauli noise present within the EV protocol.

We start with a target unitary $U(\bm{\theta})$, with $\bm{\theta}~=~(\theta_0, \theta_1, \dots)$ corresponding with angles of rotation gates, for the state preparation step in EV (see Figure \ref{fig:dsp_circ}). We may round the parameters $\bm{\theta}$ to the nearest multiples of $\frac{\pi}{2}$ to obtain a classically simulable circuit $U(\bm{\theta}^{\mathrm{clifford}}) \in \mathscr{C}(\mathscr{H})$ where
\begin{equation}
    \theta^{\mathrm{clifford}}_i = \frac{\pi}{2} \cdot \mathrm{round}\Big(\frac{2}{\pi}\cdot\theta_i\Big).
\end{equation}
The resulting entangling structure will match that of the target unitary and thus approximates the noise channel appropriately; this differs from the work of \cite{urbanek2021mitigating} where all single-qubit gates were dropped, leaving just that entangling structure in the Clifford estimation circuit. For a Pauli operator $P$ whose expectation value we wish to estimate, we will have $\mathcal{C}(\bm{\theta}^{\mathrm{clifford}}) = \bra{0}U^\dag(\bm{\theta}^{\mathrm{clifford}})PU(\bm{\theta}^{\mathrm{clifford}})\ket{0} \in \{-1,0,+1\}$. 

We may improve the quality of our Pauli noise training data by permitting a small number $L\in\mathbb{N}$ of non-Clifford gates in the training circuits, chosen such that we maintain efficiency in evaluating the relevant expectation values. In order to do so, we select at random $L$ parameters $\bm{\theta}^{\mathrm{non-clifford}} = (\theta_{i_1}, \dots, \theta_{i_L})$ to leave un-rounded in the circuit, noting that this may generate up to $2^L$ terms in the corresponding state and thus must be set such that the available classical resource may accommodate this overhead. We randomize over $M \in \mathbb{N}$ selections $\bm{i} = (i_1, \dots, i_L)$, ensuring to include only choices of parameters in the light-cone circuit of $U(\bm{\theta})$ (see Figure~\ref{fig:light-cone}) to avoid redundant data; we denote by $\mathcal{I}_{\mathrm{light-cone}}$ the indexing set of such parameters. Letting $U(\bm{\theta}; \bm{i})$ be the near-Clifford circuit whose rotation angles are rounded to the nearest multiple of $\frac{\pi}{2}$, except for those having indices $\bm{i} \subset \mathcal{I}_{\mathrm{light-cone}}$, and $\rho(\bm{\theta}, \bm{i})$ the corresponding reduced density matrix of the ancilla qubit in EV, we obtain noiseless expectation values
\begin{equation}
    \mathcal{C} = \bigg\{ \mathcal{C}(\bm{\theta}; \bm{i}) \;|\; \bm{i}\subset\mathcal{I}_{\mathrm{light-cone}} \bigg\}
\end{equation}
and $X/Z$-basis noisy measurements
\begin{equation} 
\begin{aligned}
    \mathcal{N}_X ={} & \bigg\{ \Tr{\big(X\rho(\bm{\theta},\bm{i})\big)} \;|\; \bm{i}\subset\mathcal{I}_{\mathrm{light-cone}} \bigg\}, \\
    \mathcal{N}_Z ={} & \bigg\{ \Tr{\big(Z\rho(\bm{\theta},\bm{i})\big)}  \;|\; \bm{i}\subset\mathcal{I}_{\mathrm{light-cone}} \bigg\}.
\end{aligned}
\end{equation}

In the previous Section \ref{pauli_channel} we derived a linear relationship between the Pauli error rates, defining learnable quantities $\Lambda_X, \Lambda_Z, \Omega_X$, and the noisy expectation values $\mathcal{N}_X, \mathcal{N}_Z$. By fitting the data
\begin{equation}
    \Bigg\{ \Bigg( \frac{1 - \mathcal{C}(\bm{i})^2}{1 + \mathcal{C}(\bm{i})^2}, \mathcal{N}_X(\bm{i}) \Bigg) \;|\; \bm{i}\subset\mathcal{I}_{\mathrm{light-cone}} \Bigg\}
\end{equation}
and
\begin{equation}
    \Bigg\{ \Bigg( \frac{2 \mathcal{C}(\bm{i})}{1 + \mathcal{C}(\bm{i})^2}, \mathcal{N}_Z(\bm{i}) \Bigg) \;|\; \bm{i}\subset\mathcal{I}_{\mathrm{light-cone}} \Bigg\}
\end{equation}
we obtain linear functions $f_X, f_Z$ via standard regression techniques. 

For a simple linear regression model consisting of data pairs $(x_i, y_i)$, in ordinary least-squares (OLS) one aims to find to find parameters $\alpha, \beta$ such that the residuals $r_i = y_i - \beta x_i - \alpha$ are minimized with respect to the objective function $S(\alpha, \beta) = \sum_i r_i^2$, the``sum of squared residuals" (SSR). However, OLS treats all data pairs equally, whereas it would be desirable to account for variability in the regression. Running the quantum experiments multiple times to obtain the requisite distributions is impractical, given the significant overheads already involved in such calculations; it would be convenient if we could instead leverage a single set of measurements to approximate its various population statistics. 

The idea underpinning \textit{statistical bootstrapping} is that we may resample from the observed measurement distribution to generate many additional theoretically-valid distributions that recover the noisy value in expectation. More concretely, suppose that we sample from a quantum device $n_{\mathrm{sample}}$ times and obtain the measurement outcome $m$ with frequency $f_m$. This defines a random variable $M$ with probabilities $P(M=m) = \frac{f_m}{n_{\mathrm{sample}}}$ from which we may resample without performing further quantum experiments. While this process generates no additional information, it allows us to understand how emulated fluctuations in the data affect some estimator $\mathcal{E}$ of interest; extracting resampled data sets $m_i$ with $|m_i|=n_{\mathrm{sample}}$ from $M$ allows us to obtain bootstrapped moments of $\mathcal{E}(m_i)$.

\begin{figure*}
    \centering

        \begin{subfigure}{0.8\linewidth}
        \centering
        \resizebox{0.7\linewidth}{!}{
        \begin{tikzpicture}[rotate=90]
            \node(h)[my hex select] at (3.5,{6*sqrt(3)/2}){};
            \node(h)[my hex select] at (3,{5*sqrt(3)/2}){};
            \node(h)[my hex select] at (3.5,{2*sqrt(3)/2}){};
            \node(h)[my hex select] at (3,{1*sqrt(3)/2}){};
        
            \node(h)[ancilla select] at (3.75,{4.5*sqrt(3)/2}){};
            \node(h)[ancilla select] at (1.75,{4.5*sqrt(3)/2}){};
            \node(h)[ancilla select] at (2.75,{2.5*sqrt(3)/2}){};
            \node(h)[ancilla select] at (0.75,{2.5*sqrt(3)/2}){};
        
            \node(h)[my hex select] at (1.5,{6*sqrt(3)/2}){};
            \node(h)[my hex select] at (1,{5*sqrt(3)/2}){};
            \node(h)[my hex select] at (1.5,{2*sqrt(3)/2}){};
            \node(h)[my hex select] at (1,{1*sqrt(3)/2}){};
            
          \foreach \n[count=\k] in {0,1,0,1,0,1}{
            \foreach \m in {1,2,3}{
              \node(h)[my hex] at (\n/2+\m,{\k*sqrt(3)/2}){};
              \foreach \t in {1,...,6} \node[my circ] at ($(h)+({(\t-1 )*60}:{.5})$){};
              \foreach \t in {1,...,6} \node[my circ] at ($(h)+({(\t-.5)*60}:{sqrt(1/3})$){};
            }
          }
        \end{tikzpicture}
        }
        \caption{The heavy-hex qubit topology of the devices utilized for this work. Shaded areas indicate 2-cell tiling to increase sampling throughput on the device. The circular shaded areas indicate the ancilla qubits selected to probe the spin magnetization of the adjacent site in the lattice region. For the larger 4-cell, of Figure \ref{fig:4-cell}, only two tiles/clusters are accommodated.}
        \label{fig:heavy-hex-tiling}
    \end{subfigure}
    
    \begin{subfigure}{0.49\linewidth}
        \includegraphics[width=\linewidth]{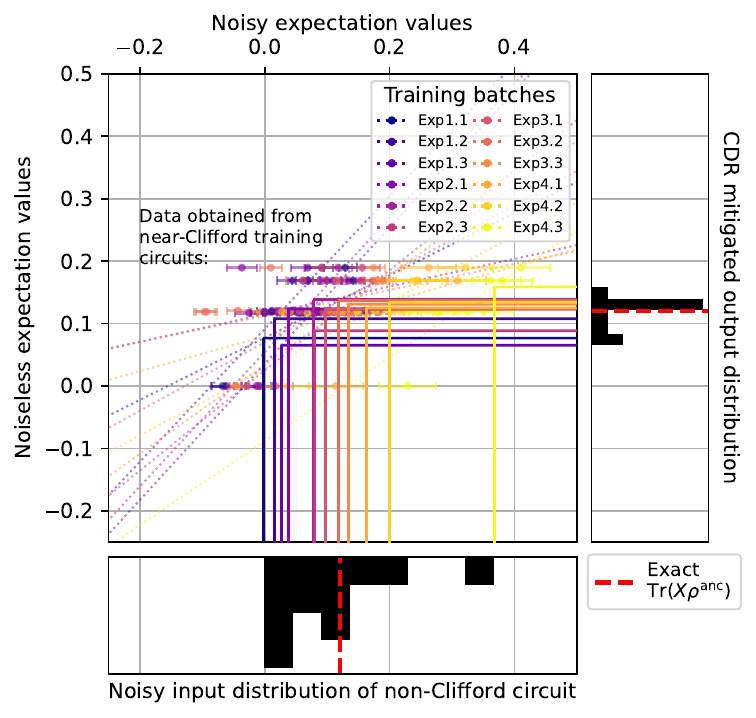}
        \caption{Estimating $\Tr(X\rho^{\mathrm{anc}})$}
    \end{subfigure}
    \begin{subfigure}{0.49\linewidth}
        \includegraphics[width=0.97\linewidth]{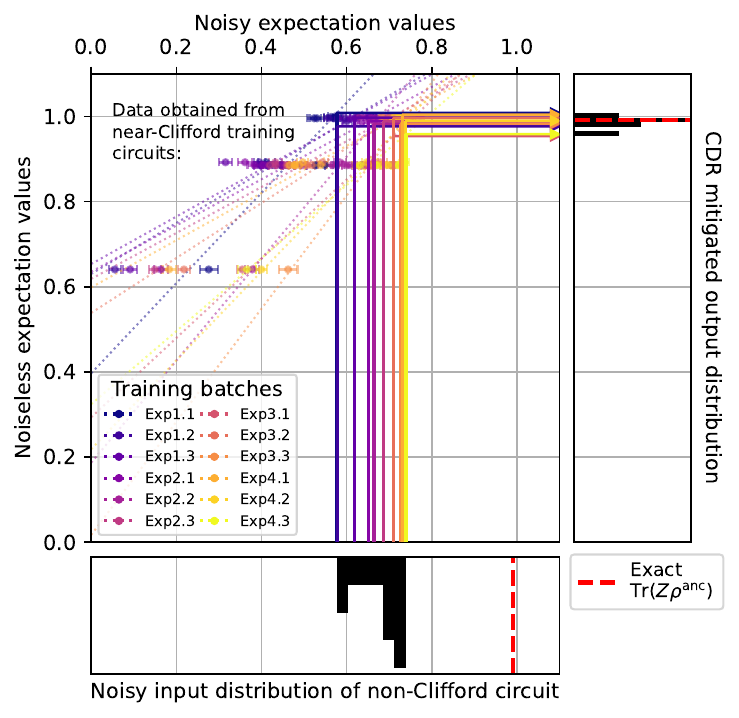}
        \caption{Estimating $\Tr(Z\rho^{\mathrm{anc}})$}
    \end{subfigure}
    
    \begin{subfigure}{\linewidth}
        \centering
        \includegraphics[width=0.6\linewidth]{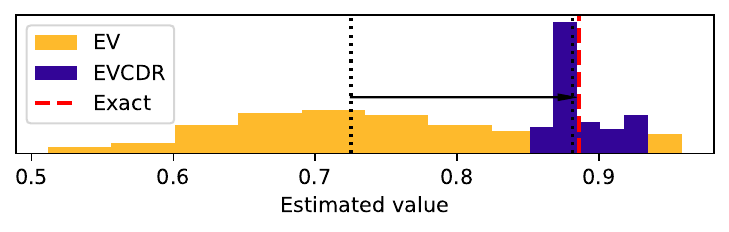}
        \caption{Estimating $\braket{V} = \Tr(Z\rho^{\mathrm{anc}})/(1+\Tr(X\rho^{\mathrm{anc}}))$}
    \end{subfigure}
    
    \caption{EVCDR for calculating the magnetization of a spin-site in a 21-qubit Ising model. \textbf{(a)} four 21-qubit clusters could be accommodated on the 127-qubit \textit{ibm\_sherbrooke} system, with three batches submitted to the Runtime service resulting in up to 12 fitting curves for near-Clifford circuit data used to learn \textbf{(b)} the noise damping and additive parameters $\Lambda_X, \Omega_X$ defined in Equation \eqref{lambdaX} and \textbf{(c)} the noise damping factor $\Lambda_Z$ defined in Equation \ref{lambdaZ}.
    Combining the ancilla expectation values in \textbf{(b)} and \textbf{(c)} yields \textbf{(d)} the final magnetization estimate as per Equation \eqref{EV_clifford_estimator_interpolated}. A maximum of 15 non-Clifford gates were permitted in each training circuit; non-Clifford rotations were selected from the light-cone at random, while the remaining angles were rounded to the nearest multiple of $\frac{\pi}{2}$. Different colours in \textbf{(b)}/\textbf{(c)} indicate training batches/clusters and arrows reflect noisy data from the true, non-rounded circuit through the Clifford curves fitted to five unique Clifford realizations per batch, yielding the final error-mitigated expectation values. Histograms indicate the estimated expectation values before and after CDR mitigation with the dashed red lines the noiseless values.}
    \label{fig:clifford_regression}
\end{figure*}

We applied this bootstrapping method in previous work to improve the extrapolation procedure in ZNE \cite{weaving2023benchmarking}. Here, it allows us to make use of weighted least-squares (WLS) regression in the Clifford learning phase of the EVCDR algorithm, which biases the fitting procedure towards data of low variance. This is achieved by defining weights $W_i$ and modifying the objective function $S(\alpha, \beta) = \sum_i W_i r_i^2$; taking $W_i = \frac{1}{\sigma_i^2}$ where $\sigma_i^2$ is the variance of $y_i$ it can be proved this approach yields a best linear unbiased estimator \cite{aitken1936iv}.

Recalling the noisy estimators $\mathcal{E}_Z, \mathcal{E}_X$ that define the standard EV estimator \eqref{EV_standard_estimator}, we may invert the Clifford fitting curves to obtain mitigated estimators $f_Z^{-1}(\mathcal{E}_Z), f_X^{-1}(\mathcal{E}_X)$ and subsequently our Echo Verified Clifford Data Regression (EVCDR) estimator
\begin{equation}\label{EV_clifford_estimator_interpolated}
    \mathcal{E}_{\mathrm{EVCDR}} = f_Z^{-1}(\mathcal{E}_Z) \Big( 1+f_X^{-1}(\mathcal{E}_X) \Big)^{-1}.
\end{equation}

In Figure \ref{fig:clifford_regression} we observe the intended effect of this process on the \textit{ibm\_sherbrooke} system, where we execute concurrent EVCDR procedures across numerous qubit clusters on the chip in batches. While there is significant variability in the noisy data obtained from $\mathcal{E}_X$ and $\mathcal{E}_Z$, when reflected through the Clifford fitting curves the estimates $f^{-1}_X(\mathcal{E}_X)$ and $f^{-1}_Z(\mathcal{E}_Z)$ become focused and variance is suppressed. Indeed, in performing calculations over restricted qubit subsets we find that it is possible to invert the noise channel locally.

\section{Software and Hardware Implementation Details}\label{sec:methods}

The following benchmark of EVCDR was deployed on the 127-qubit Eagle r3 device \textit{ibm\_sherbrooke}\hl{, with the native gateset $\{X, \sqrt{X}, R_z, \mathrm{ECR}\}$} and a rated error per layered gate (EPLG) of $1.7\%$ at the time of execution. For each simulation we tiled as many concurrent circuit instances across the chip as possible in order to passively average hardware errors and increase the effective yield from the $80,000$ circuit shots performed at each time step; the heavy-hex 2-cell permitted four circuit tilings (example in Figure \ref{fig:heavy-hex-tiling}), while the 4-cell allowed just two. We performed CDR within each cluster to learn the effect of the local noise channel, as opposed to pooling the data into a single CDR routine; this is visible in Figure \ref{fig:clifford_regression}, where each fitting line represents different clusters/batches to account for the non-uniformity of device noise. Note that each CDR circuit received the same shot allowance across three batches of five unique Clifford realizations, resulting in fifteen separate backend submissions in addition to the standard EV one. This could be argued as an unfair advantage, whereas the budget should have been distributed across each training circuit.

To mitigate the problem of vanishing success probability discussed in Section \ref{sec:EV_depolarizing} and Figure \ref{fig:p0}, we allow a small number of bit-flips in the postselection procedure; we implement this by forcing a neighbourhood of qubits around the ancilla to be zero, while those outside the selected neighbourhood may deviate by a predefined maximum Hamming distance. Although we may now be including invalid measurements that will bias our output, we argue that it is a reasonable assumption as a small number of flips will correspond with a small change to the magnetization, particularly if the flips occur far from the measured spin-site; see Appendix \ref{sec:volumetric}. 

We developed a bespoke implementation for the circuit construction, clustering, batching and final submission to the quantum backend on top of \textit{Qiskit} functionality \cite{Qiskit}. Pauli operator manipulation and Clifford simulation was handled using the \textit{Symmer} package \cite{symmer2022} and the regression step in CDR made use of \textit{statsmodels} \cite{seabold2010statsmodels}. We have provided all the code and data on an online repository \cite{EVCDR_repo} so that readers may reproduce the results presented here, or design their own EV-based error mitigation workflows.

\section{Results}\label{sec:results}

\begin{figure*}
    \centering
    \input{light_cone_topology_expansion_4_cell}
    \caption{The heavy-hex 4-cell, consisting of 35 qubits. Figures \textbf{(a)}-\textbf{(g)} illustrate the expansion of the light-cone for successive time steps $K$ probing time $t=K\tau$, until the spin lattice is saturated from $K=7$ onwards. Different colours indicate the additional qubits included in the light-cone at each step}.
    \label{fig:4-cell}
\end{figure*}

To benchmark our EVCDR method we choose to simulate the Ising model on 2D spin-lattices
\begin{equation}\label{H_ising}
    H = - J \sum_{\braket{a,b}} Z^{(a)} Z^{(b)} - h \sum_a X^{(a)}
\end{equation}
where $J,h \in \mathbb{R}$ parameterize strengths of the spin coupling and external field, respectively; the ratio $J/h$ dictates the Hamiltonian time evolution. The pairwise site indices $\braket{a,b}$ run over nodes of the graph representing the nearest-neighbour spin couplings, which we choose such that the corresponding lattice is subgraph isomorphic with respect to the hardware topology; see the heavy-hex 4-cell in Figure \ref{fig:4-cell}. 

The quantity we aim to estimate is the time-dependent magnetization on a single spin-site
\begin{equation}
    M(t) = \bra{0} e^{-iHt} Z^{(0)} e^{iHt} \ket{0}.
\end{equation}
For a number of time steps $K \in \mathbb{N}$ with step-size $\tau$, the Hamiltonian evolution at final time $t = K\tau$ is captured by a circuit $\prod_{k=1}^{K} U(\tau)$ where
\begin{equation}
\begin{aligned}
    U(\tau) ={} & \prod_{\braket{a,b}} e^{-iJ\tau Z^{(a)}Z^{(b)}} \prod_{a} e^{-ih\tau X^{(a)}} \\
    ={} & e^{iH\tau} + \mathcal{O}(\tau^2).
\end{aligned}
\end{equation}

The circuit formulation of $U(\tau)$ is depicted in Figure \ref{fig:ising_circuit}, which propagates the system through a time $\tau$ with each succesive application. The error here is the consequence of first-order Trotterization \cite{childs2021theory} and may be combated by taking finer time slices~$\tau$, with the trade-off being increased circuit depth. With each successive application of the time propagation circuit, the effective light-cone for local observables expands as illustrated in Figure \ref{fig:light-cone}. In Figure \ref{fig:4-cell} we have indicated how this expansion grows from a single spin-site on the heavy-hex $4$-cell lattice with each successive time step $K$ for our Ising model time evolution problem. From $K=7$ onward the lattice is saturated.

For this practical demonstration we compare the following echo-verified estimators:
\begin{itemize}
    \item \textbf{Standard}: the approach outlined in the introductory Section \ref{sec:intro}.
    \item \textbf{Spectral Purification}: a modification to the standard estimator, whereby one decomposes the state of the ancilla qubit $\rho^{\mathrm{anc}} = \lambda_0 \ket{\phi_0}\bra{\phi_0} + \lambda_1 \ket{\phi_1}\bra{\phi_1}$ and selects the dominant pure component, say $\lambda_0>\lambda_1$ (discussed in \cite{huo2022dual, weaving2023benchmarking}).
    \item \textbf{Purity Normalization}: through the observation $\gamma(\rho^{\mathrm{anc}})^2 \approx \Tr{(X\rho^{\mathrm{anc}})}^2 + \Tr{(Y\rho^{\mathrm{anc}})}^2 + \Tr{(Z\rho^{\mathrm{anc}})}^2$ where $\gamma(\rho^{\mathrm{anc}})$ is the purity, one may re-normalize the ancillary estimators $\mathcal{E}_X \mapsto \mathcal{E}_X/\gamma(\rho^{\mathrm{anc}}), \mathcal{E}_Z \mapsto \mathcal{E}_Z/\gamma(\rho^{\mathrm{anc}})$ to recover $\mathcal{E}_X^2 + \mathcal{E}_Z^2 = 1$, the property used to form alternative EV estimators \eqref{EV_variant}.
    \item \textbf{Clifford Data Regression}: approximate the target unitary with near-Clifford circuits that may be evaluated classically to learn the effect of noise on the EV estimate, as detailed in Section \ref{clifford_learning} and pictured in Figure \ref{fig:EV_flowchart}. The noise manifests as a linear relationship between the noisy and ideal expectation values and can therefore be analysed via regression techniques.
\end{itemize}

\begin{figure}[t!]
    \centering
    \resizebox{0.85\linewidth}{!}{\input{ising_circuit}}
    \caption{A single time step circuit for simulating the Ising Hamiltonian \eqref{H_ising}. Evolution for longer times is achieved via repeated application of this circuit block. The rotation convention is $R_x(\theta) = \exp{(-i\frac{\theta}{2} X)}$ and $R_{z}(\theta) = \exp{(-i\frac{\theta}{2} Z)}$.}
    \label{fig:ising_circuit}
\end{figure}

\begin{figure}[h]
    \centering
    \begin{subfigure}{\linewidth}
    \resizebox{0.95\linewidth}{!}{
    \begin{quantikz}[row sep=0.6cm, column sep=0.2cm, wire types={q,q}]
     & \ctrl{1} & \\
     & \targ{}  &
    \end{quantikz}
    $\propto$
    \begin{quantikz}[row sep=0.2cm, column sep=0.21cm, wire types={q,q}]
    &                  &                 & \gate{R_z(-\frac{\pi}{2})} & \gate[2]{\mathrm{ECR}} & \gate{X} & \\
    & \gate{R_z(-\pi)} & \gate{\sqrt{X}} & \gate{R_z(-\pi)}           &                        &          &
    \end{quantikz}
    }
    \caption{Global phase $\frac{\pi}{2}$}
    \end{subfigure}
    \vspace{0mm}
    
    \begin{subfigure}{\linewidth}
    \resizebox{0.95\linewidth}{!}{
    \begin{quantikz}[row sep=0.6cm, column sep=0.2cm, wire types={q}]
     & \gate{R_x(2h\tau)} &
    \end{quantikz}
    $\propto$
    \begin{quantikz}[row sep=0.2cm, column sep=0.2cm, wire types={q}]
    & \gate{R_z(\frac{\pi}{2})}&\gate{\sqrt{X}}&\gate{R_z(2h\tau+\pi)}&\gate{\sqrt{X}}& \gate{R_z(\frac{5\pi}{2})} & 
    \end{quantikz}
    }
    \caption{Global phase $\frac{3\pi}{2}$}
    \end{subfigure}
    \caption{\hl{Transpilation of non-native gates in Figure \ref{fig:ising_circuit} onto the target gateset $\{X, \sqrt{X}, R_z, \mathrm{ECR}\}$.}}
    \label{fig:test}
\end{figure}

\begin{figure*}
\centering
\begin{subfigure}{0.484\linewidth}
    \centering
    % \hspace{1.2cm}
    % \textbf{Heavy-hex 2-cell}
    \begin{tikzpicture}[rotate=90]
    \node(h1)[my hex] at (2,{1*sqrt(3)/2}){};
    \node(h2)[my hex] at (1,{1*sqrt(3)/2}){};
    \foreach \t in {1,...,6} \node[big circ] at ($(h2)+({(\t-.5)*60}:{sqrt(1/3})$){};
    \foreach \t in {1,...,6} \node[big circ] at ($(h2)+({(\t-1 )*60}:{.5})$){};
    \foreach \t in {1,...,6} \node[big circ] at ($(h1)+({(\t-1 )*60}:{.5})$){};
    \foreach \t in {1,...,6} \node[big circ] at ($(h1)+({(\t-.5)*60}:{sqrt(1/3})$){};
    \node at ($(h1)+({(5-.5)*60}:{1.3})$){$\rightarrow\braket{Z}$};
    \node[big circ, fill=black] at ($(h1)+({(5-.5)*60}:{sqrt(1/3})$){};
    \end{tikzpicture}
    \includegraphics[width=\linewidth]{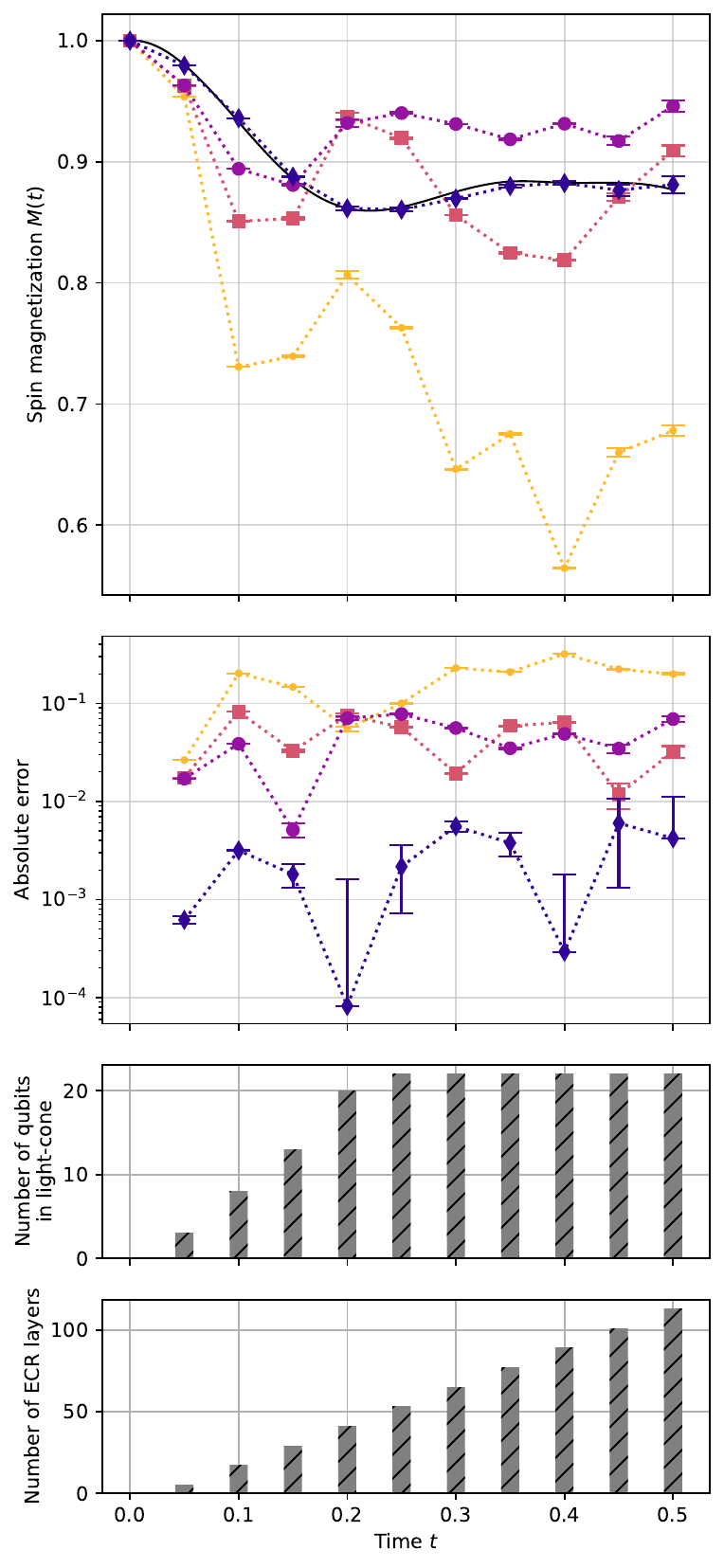}
    \caption{}
    \label{fig:evolution_2_cell}
\end{subfigure}
\begin{subfigure}{0.496\linewidth}
    \centering
    % \hspace{1.2cm}
    % \textbf{Heavy-hex 4-cell}
    \begin{tikzpicture}[rotate=90]
    \node(h1)[my hex] at (2,{1*sqrt(3)/2}){};
    \node(h2)[my hex] at (1,{1*sqrt(3)/2}){};
    \node(h3)[my hex] at (1+0.5,{2*sqrt(3)/2}){};
    \node(h4)[my hex] at (1+0.5,{0*sqrt(3)/2}){};
    \foreach \t in {1,...,6} \node[big circ] at ($(h1)+({(\t-1 )*60}:{.5})$){};
    \foreach \t in {1,...,6} \node[big circ] at ($(h2)+({(\t-.5)*60}:{sqrt(1/3})$){};
    \foreach \t in {1,...,6} \node[big circ] at ($(h2)+({(\t-1 )*60}:{.5})$){};
    \foreach \t in {1,...,6} \node[big circ]
    at ($(h1)+({(\t-.5)*60}:{sqrt(1/3})$){};
    \foreach \t in {1,...,6} \node[big circ] at ($(h3)+({(\t-1 )*60}:{.5})$){};
    \foreach \t in {1,...,6} \node[big circ] at ($(h4)+({(\t-.5)*60}:{sqrt(1/3})$){};
    \foreach \t in {1,...,6} \node[big circ] at ($(h4)+({(\t-1 )*60}:{.5})$){};
    \foreach \t in {1,...,6} \node[big circ]
    at ($(h3)+({(\t-.5)*60}:{sqrt(1/3})$){};
    \node at ($(h4)+({(5-.5)*60}:{1.3})$){$\rightarrow\braket{Z}$};
    \node[big circ, fill=black] at ($(h4)+({(5-.5)*60}:{sqrt(1/3})$){};
    \end{tikzpicture}
    \includegraphics[width=\linewidth]{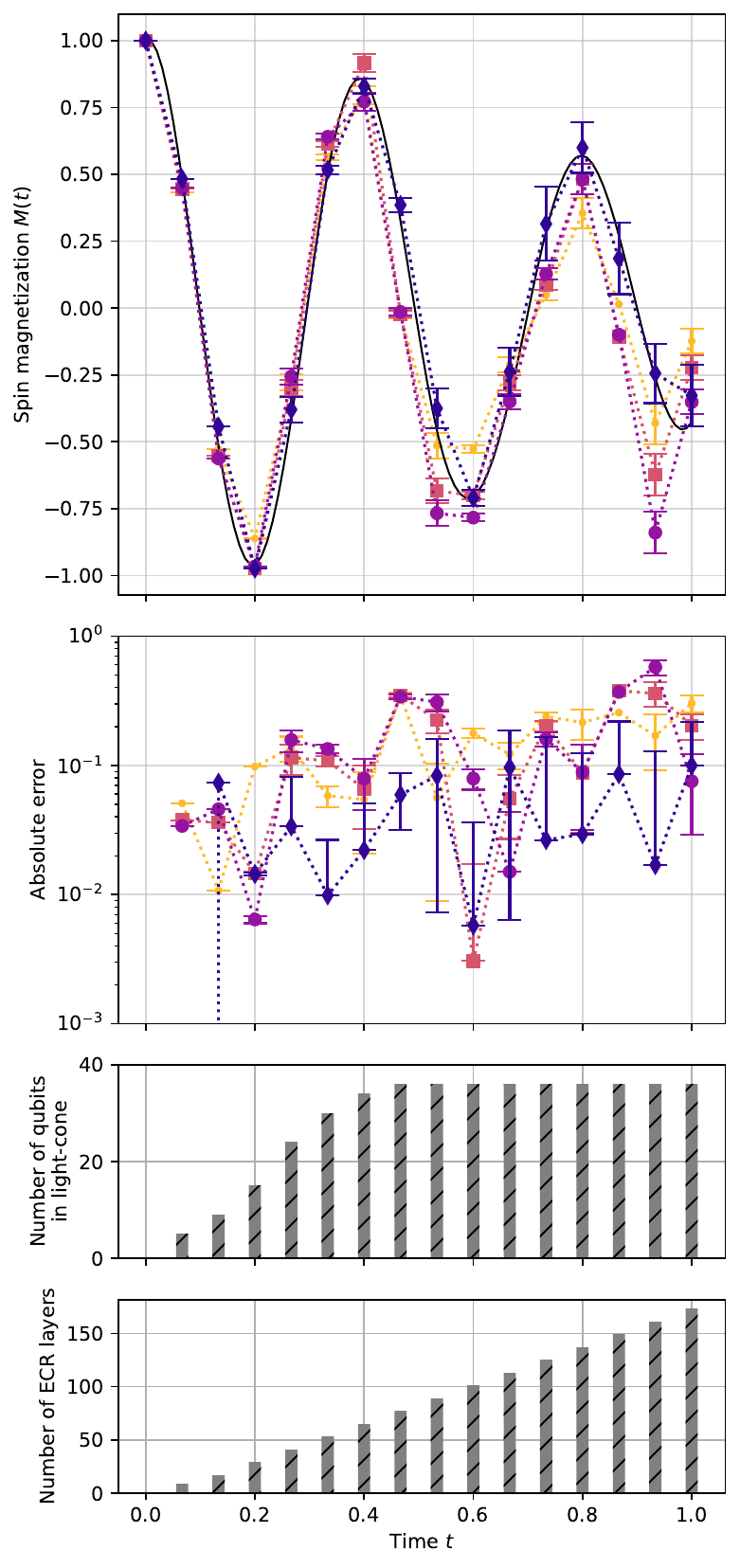}
    \caption{}
    \label{fig:evolution_4_cell}
\end{subfigure}
\includegraphics[width=\linewidth]{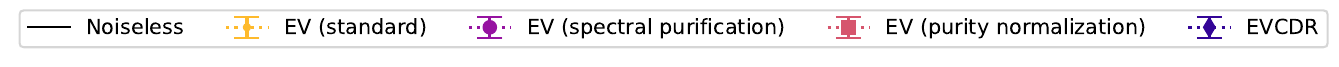}
\caption{Comparison of Echo Verification estimators to calculate the spin magnetization evolution for a single site in heavy-hex Ising models \eqref{H_ising}. \textbf{(a)} Heavy-hex 2-cell (21-qubits) Ising model with $J=4, h=2$, evolved to a final time $T=1/2$ over $10$ steps, corresponding with a step size of $\tau = 1/20$. \textbf{(b)} Heavy-hex 4-cell (35-qubits) Ising model with $J=1, h=8$, evolved to a final time $T=1$ over $15$ steps, corresponding with a step size of $\tau = 1/15$. Error bars are obtained from statistical bootstrapping of the experimental measurement distributions, described in Section \ref{clifford_learning}, and show standard error on the mean. Note the errors are approximately an order of magnitude lower for the $2$-cell compared with the $4$-cell.}

\end{figure*}

We first present the heavy-hex 2-cell results of Figure \ref{fig:evolution_2_cell}, consisting of 21-qubits in total. With respect to the Ising model \eqref{H_ising} with $J=4, h=2$, we observe the evolution of the magnetization of a single spin over $10$ steps to a final time $T=1/2$ in increments of $\tau = 1/20$. Due to the light-cone reduction depicted in Figure \ref{fig:light-cone}, we do not saturate the spin lattice until time step $K=5$, while the \hl{number of entangling ECR layers continues to increase linearly past this point to a maximum of $113$ at $K=10$}. We find that the standard EV approach yields poor errors in excess of $0.1$, while the purity normalized and spectral purification modifications suppress the absolute error to within $0.1$ and $0.01$. EVCDR, on the other hand, consistently produces the lowest errors in the range $0.01 - 0.0001$.

Following the 21-qubit case, we then attempt the EVCDR technique on the heavy-hex 4-cell defined over 35-qubits visible in Figure \ref{fig:evolution_4_cell}. In the Ising model we take $J=1, h=8$ and evaluate the evolution to a final time of $T=1$ in 15 steps of size $\tau=1/15$. Here, the light-cone does not saturate the lattice until time step $K=7$, with the \hl{number of entangling ECR layers reaching a maximum of $173$ at the final step $K=15$}. Once again, EVCDR offers the most consistent results with absolute errors never exceeding 0.1 throughout the time evolution, with many points dropping to order 0.01 in error. Visually comparing the results in Figures \ref{fig:evolution_2_cell} and \ref{fig:evolution_4_cell}, it might appear that all the EV methods investigated here perform better for the larger 35-qubit simulation; however, we draw attention to the scale on the Y-axis where the 4-cell evolves between $[-1,+1]$ while the 2-cell is contained in $[0.85,1]$. The error plot is more instructive, indicating approximately one order of magnitude increase in error going from the 2-cell to the larger 4-cell.

\section{Discussion}

In this work we performed a rigorous analysis of the noise present within the Echo Verification quantum error mitigation technique. We discovered that the presence of noise results in a linear relationship between the noisy and ideal expectation values, which we characterized explicitly for a simplified model of Pauli noise, which assumed all noise may be delayed to the end of the circuit and applied to the final density operator. This holds strictly for Clifford circuits only, so future work may look to expand the noise analysis to a more general model.

These findings motivated the application of a Clifford Data Regression routine, through which we may learn the effect of the noise channel via near-Clifford training circuits since we may efficiently evaluate the circuits classically and compare with the noisy results extracted from hardware. We found the resulting quantum error mitigation methodology, Echo Verified Clifford Data Regression (EVCDR), to be robust against noise, as exemplified by successfully evaluating the time evolution for Ising models defined over spin lattices of 21 and 35 qubits in size on superconducting quantum hardware. 

While EVCDR is effective for these systems, there still remains some concern over the sampling overhead due to the postselection procedure encountered in EV-based techniques. Since EV-based methods are in general shot-bound, one could argue the CDR addition worsens this scaling obstacle, despite its efficacy. Furthermore, it is known that expectation values of random (near) Clifford circuits tend to concentrate around zero for sufficiently large systems, which research has tried to combat through ``peaked" circuits \cite{aaronson2024verifiable}. While this was not a limitation for our simulations, it should be taken into consideration when scaling EV. Alongside the success probability question, future work should examine these issues closely in the pursuit of applying Echo Verification methods to larger systems.

\section*{Acknowledgements}
T.W. is funded by EPSRC (EP/S021582/1) and CBKSciCon Ltd.. A.R. and P.J.L. acknowledge support by the NSF STAQ project (PHY-1818914). S.S. wishes to acknowledge financial support from the National Centre for HPC, Big Data and Quantum Computing (Spoke 10, CN00000013). P.V.C. is grateful for funding from the European Commission for VECMA (800925) and EPSRC for SEAVEA (EP/W007711/1). Access to hardware was facilitated by the Italian Institute of Technology (IIT) through the IBM Quantum Hub at CERN.

\bibliographystyle{apsrev4-2_mod.bst}
\bibliography{main}

\clearpage
\onecolumngrid
\appendix

\section{Volumetric Benchmarking of Loschmidt Echos}\label{sec:volumetric}

\begin{figure}[b!]
    \centering
    \begin{subfigure}{0.44\linewidth} 
        \centering
        \includegraphics[width=\linewidth]{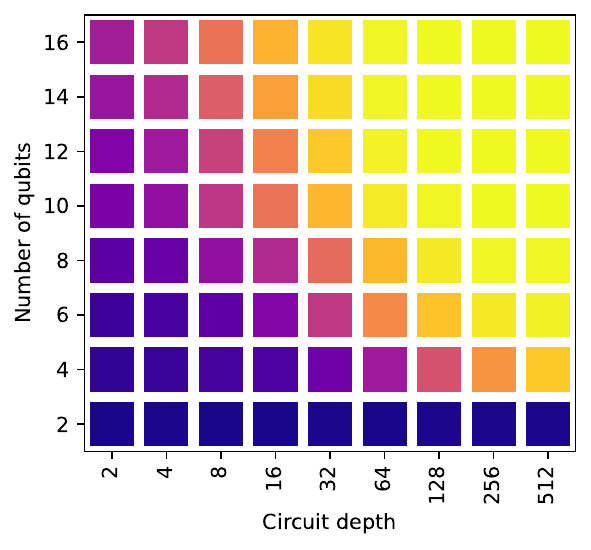}
        \caption{}
        \label{fig:volume_raw}
    \end{subfigure}
    \begin{subfigure}{0.44\linewidth} 
        \centering
        \includegraphics[width=\linewidth]{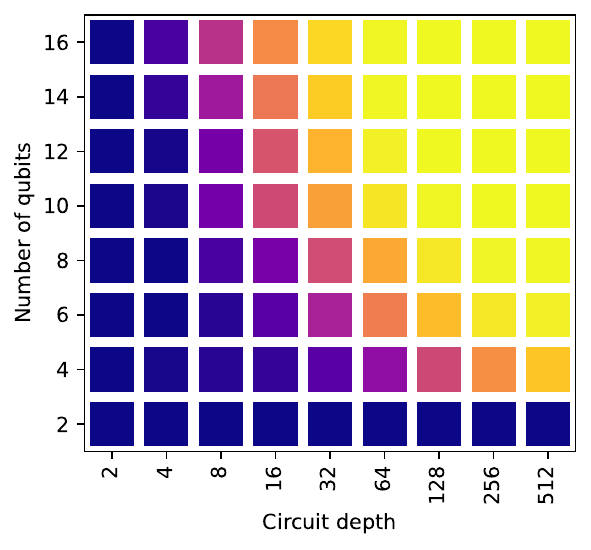}
        \caption{}
        \label{fig:volume_mem}
    \end{subfigure}
    \begin{subfigure}{0.1\linewidth}
        \centering
        \includegraphics[width=0.9\linewidth]{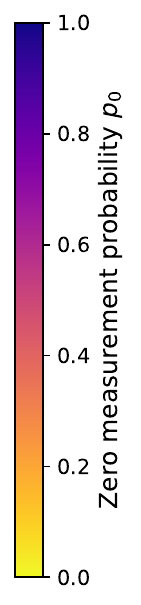}
        \vspace{8mm}
    \end{subfigure}
    \caption{Volumetric benchmarking of the zero measurement probability $p_{\bm{0}}$ for Loschmidt echos $UU^\dag$ over random samples of unitaries $U$ on the \textit{ibm\_strasbourg} device. The experiments each used $8,192$ shots and utilized Dynamical Decoupling and Pauli Twirling, either \textbf{(a)} \textit{without} or \textbf{(b)} \textit{with} Measurement-Error Mitigation.}
    \label{fig:volumetric_benchmark}
\end{figure}

As discussed in Section \ref{sec:EV_depolarizing}, the zero postselection probability is one of the greatest obstacles to scaling the EV method. In Equation \eqref{p0_depolarizing} we found this probability to be 
\begin{equation}
    p_{\bm{0}}(\delta) = \frac{1+\braket{V}^2}{2} \cdot (1-\delta)+\frac{2\delta}{d}
\end{equation}
where $V$ is the unitary we are measuring and $\delta$ is the depolarization rate. While in the noiseless case we have $p_{\bm{0}}(0) \geq 0.5$, the presence of noise can suppress the probability arbitrarily close to zero as $\delta \rightarrow 1$. This is affected by many factors, such as gate noise and readout error. To combat this, other error mitigation/suppression techniques such as Dynamical Decoupling (DD), Pauli Twirling (PT) and Measurement-Error Mitigation (MEM) are compatible with EV/EVCDR.

In this section we perform a volumetric benchmark \cite{blume2020volumetric}  of the zero measurement probability for Loschmidt echos $UU^\dag$ where the unitaries $U$ are generated uniformly at random. We note the Loschmidt echo is closely related to EV, as it corresponds to the trivial case $V=I$ in which case we expect to find $p_{\bm{0}} = 1$. In particular, we probe the effect of MEM on top of both DD and PT to see by how much we may amplify the postselection probability in Figure \ref{fig:volumetric_benchmark}.

\begin{figure}[t!]
    \centering
    \includegraphics[width=0.85\linewidth]{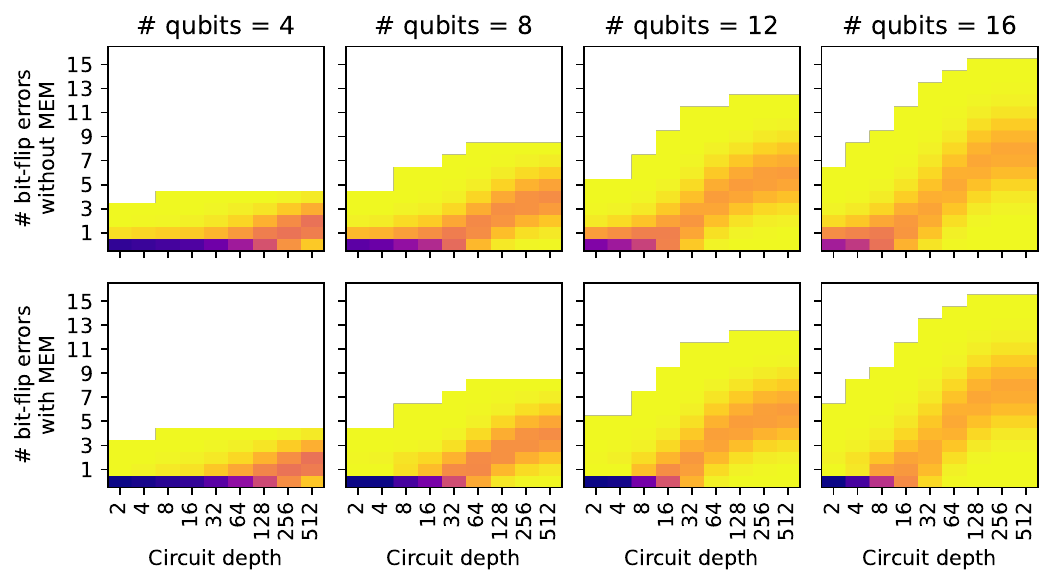}
    \caption{Observed probability of bit-flip errors for Loschmidt echos, from $8,192$ shots on \textit{ibm\_strasbourg}.}
    \label{fig:bit_flip_errors}
\end{figure}

\begin{figure}[t!]
    \centering
    \begin{subfigure}{0.53\linewidth} 
    \resizebox{\linewidth}{!}{
    \begin{tikzpicture}[rotate=90]
     \node(h1)[my hex,minimum size=1.5cm, fill=plasma6!40, draw=none] at (2,{1*sqrt(3)/2}){};
    \node(h2)[my hex,minimum size=1.5cm, fill=plasma6!40,draw=none] at (1,{1*sqrt(3)/2}){};
    \node(h3)[my hex,minimum size=1.5cm, fill=plasma6!40,draw=none] at (1+0.5,{2*sqrt(3)/2}){};
    \node(h4)[my hex,minimum size=1.5cm, fill=plasma4!80,draw=none] at (1+0.5,{0*sqrt(3)/2}){};
    \node[rectangle, fill=plasma6!40, minimum width=1cm, minimum height = 1.31cm] at (1.5,.36){};
    \node[text=plasma4,align=center,minimum size=1cm] at (2.5,-1){\tiny No bit-flips permitted};
    \node[text=plasma6,align=center,minimum size=1cm] at (0,0.2){\tiny Small number of bit-flips permitted};
    \node(h1)[my hex] at (2,{1*sqrt(3)/2}){};
    \node(h2)[my hex] at (1,{1*sqrt(3)/2}){};
    \node(h3)[my hex] at (1+0.5,{2*sqrt(3)/2}){};
    \node(h4)[my hex] at (1+0.5,{0*sqrt(3)/2}){};
    \foreach \t in {1,...,6} \node[big circ, fill=plasma6] at ($(h1)+({(\t-1 )*60}:{.5})$){};
    \foreach \t in {1,...,6} \node[big circ, fill=plasma6] at ($(h2)+({(\t-.5)*60}:{sqrt(1/3})$){};
    \foreach \t in {1,...,6} \node[big circ, fill=plasma6] at ($(h2)+({(\t-1 )*60}:{.5})$){};
    \foreach \t in {1,...,6} \node[big circ, fill=plasma6]
    at ($(h1)+({(\t-.5)*60}:{sqrt(1/3})$){};
    \foreach \t in {1,...,6} \node[big circ, fill=plasma6] at ($(h3)+({(\t-1 )*60}:{.5})$){};
    \foreach \t in {1,...,6} \node[big circ, fill=plasma6] at ($(h4)+({(\t-.5)*60}:{sqrt(1/3})$){};
    \foreach \t in {1,...,6} \node[big circ, fill=plasma6] at ($(h4)+({(\t-1 )*60}:{.5})$){};
    \foreach \t in {1,...,6} \node[big circ, fill=plasma6]
    at ($(h3)+({(\t-.5)*60}:{sqrt(1/3})$){};
    \node at ($(h4)+({(5-.5)*60}:{1.3})$){$\rightarrow\braket{Z}$};
    \node[big circ, fill=black] at ($(h4)+({(5-.5)*60}:{sqrt(1/3})$){};
    \node[big circ, fill=plasma4] at ($(h4)+({(4-.5)*60}:{sqrt(1/3})$){};
    \node[big circ, fill=plasma4] at ($(h4)+({(6-.5)*60}:{sqrt(1/3})$){};
    \node[big circ, fill=plasma4] at ($(h4)+({(5-1 )*60}:{.5})$){};
    \node[big circ, fill=plasma4] at ($(h4)+({(6-1 )*60}:{.5})$){};
    \end{tikzpicture}
    }
    \caption{}
    \end{subfigure}
    \begin{subfigure}{0.46\linewidth} 
    \centering
    \includegraphics[width=\linewidth]{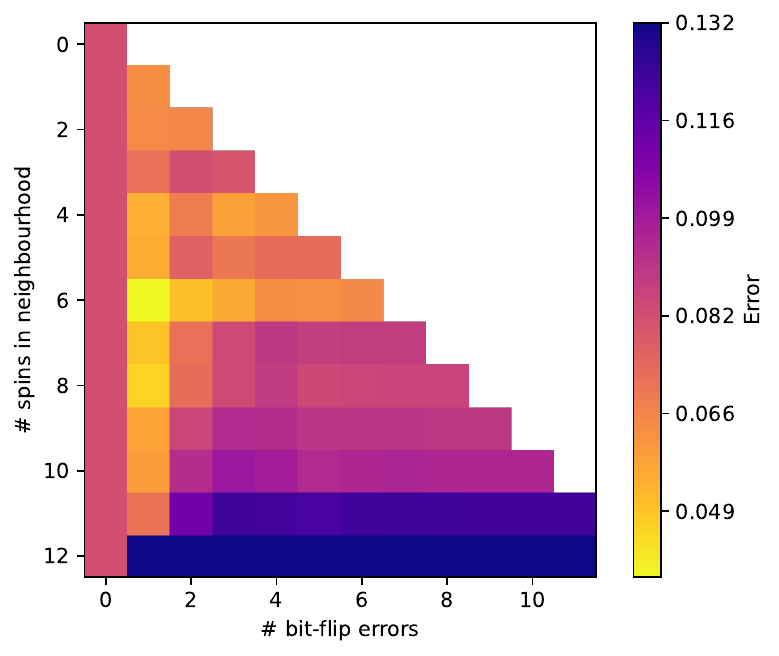}
    \caption{}
    \end{subfigure}
    \caption{\textbf{(a)} To boost the postselection probability in our implementation of Echo Verification we permitted a small number of bit-flip errors outside a neighbourhood of spin-sites around the readout qubit whose magnetization we evaluated in Figures \ref{fig:evolution_2_cell} and \ref{fig:evolution_4_cell}. For the heavy-hex 4-cell lattice we took the neighbourhood to be the four sites surrounding the readout qubit, while the remaining 30 qubits were permitted up to four bit-flip errors in the postselection procedure. \textbf{(b)} EV error for a 12-qubit Ising ring obtained for different sized qubit neighbourhoods and permitted number of bit-flip errors, simulated using the \textit{FakeGuadalupeV2} backend.}
    \label{fig:neighbourhood_qubits}
\end{figure}

We find for shallow circuits, where we expect the readout error to be the greatest contributor to the degradation of the measurement results, that MEM is very successful. However, for deep circuits it does not appreciably improve the postselection probability. Moreover, in Figure \ref{fig:bit_flip_errors} we observe the distribution of bit-flip errors and find here that MEM does not significantly affect the results, again due to the bit-flip effects being prevalent for deeper circuits.

Figure \ref{fig:bit_flip_errors} reveals a concentration of bit-flip errors, which accumulates binomially with the circuit depth, while small numbers of erroneous bit-flips occur with low probability. In an attempt to boost the postselection probability in Section \ref{sec:results}, we permit a small number of bit-flip errors outside a neighbourhood of qubits surrounding the spin-site whose magnetization we probe in the simulations of Figure \ref{fig:evolution_2_cell} and \ref{fig:evolution_4_cell}. In Figure \ref{fig:neighbourhood_qubits} we illustrate the actual neighbourhood used for the heavy-hex 4-cell lattice, indicated by the four spin-sites surrounding the readout qubit. We permit at most four bit-flip errors across the remaining 30 spins, which boosts the postselection probability sufficiently to obtain the results of Figure \ref{fig:evolution_4_cell}. We treat the 2-cell lattice similarly.

\section{Ancilla State Under Arbitrary Pauli Noise}\label{sec:arb_pauli_noise}

We explicitly partition the qubits into the system $\mathscr{H}^{(\mathrm{sys})} = (\mathbb{C}^2)^{\otimes N}$ and ancilla $\mathscr{H}^{(\mathrm{anc})} = \mathbb{C}^2$ registers so that $\mathscr{H} = \mathscr{H}^{(\mathrm{sys})} \otimes \mathscr{H}^{(\mathrm{anc})}$. We may write a Pauli $P \in \mathscr{B}(\mathscr{H})$ across this division $P = P^{(\mathrm{sys})} \otimes P^{(\mathrm{anc})}$ where $P^{(\mathrm{sys})} \in \{I,X,Y,Z\}^{\otimes N}$ and $P^{(\mathrm{anc})}~\in~\{I,X,Y,Z\}$ acts on just a single qubit. We associate with each Pauli operator $P_i \in \mathscr{B}(\mathscr{H})$ a probability $\lambda_i$ that the error $\rho \mapsto P_i \rho P_i$ occurs; the corresponding noise channel therefore assumes the form $\Phi(\rho) \coloneqq \sum_{i} \lambda_i P_i \rho P_i$. In Echo Verification we are interested in the noiseless state 
\begin{equation}
\rho = \ket{\psi}\bra{\psi}\;\; \text{where}\;\; \ket{\psi} = \frac{1}{\sqrt{2}}\Big(\ket{\bm{0}}^{(\mathrm{sys})}\otimes\ket{+}^{(\mathrm{anc})}+U^\dag V U \ket{\bm{0}}^{(\mathrm{sys})}\otimes\ket{-}^{(\mathrm{anc})}\Big),
\end{equation}
as per Equation \eqref{standard_ev_psi}. Letting $M_0 = \ket{\bm{0}}\bra{\bm{0}} \otimes I$ be the projector onto the zero state of $\mathscr{H}^{(\mathrm{sys})}$, we wish to evaluate $M_0 \Phi(\rho) M_0$ in order to understand how noise propagates from the system through to the ancilla qubit. There are four terms, labelled below $\circled{1}-\circled{4}$ so they may be addressed separately, that we need to consider in the expansion of $\Phi(\rho)$ for each index $i$:
\begin{equation}\label{full_phi_rho}
\begin{aligned}
    \Phi(\rho) ={} & \sum_i \lambda_i \bigg[ \\
    \circled{1} \leftarrow & P_i^{(\mathrm{sys})}\ket{\bm{0}}\bra{\bm{0}}P_i^{(\mathrm{sys})} \otimes P_i^{(\mathrm{anc})}\ket{+}\bra{+}P_i^{(\mathrm{anc})} + \\
    \circled{2} \leftarrow & P_i^{(\mathrm{sys})} U^\dag V U \ket{\bm{0}}\bra{\bm{0}} U^\dag V U P_i^{(\mathrm{sys})} \otimes P_i^{(\mathrm{anc})}\ket{-}\bra{-}P_i^{(\mathrm{anc})} + \\
    \circled{3} \leftarrow & P_i^{(\mathrm{sys})} \ket{\bm{0}}\bra{\bm{0}} U^\dag V U P_i^{(\mathrm{sys})} \otimes P_i^{(\mathrm{anc})}\ket{+}\bra{-}P_i^{(\mathrm{anc})} + \\
    \circled{4} \leftarrow & P_i^{(\mathrm{sys})} U^\dag V U \ket{\bm{0}}\bra{\bm{0}} P_i^{(\mathrm{sys})} \otimes P_i^{(\mathrm{anc})}\ket{-}\bra{+}P_i^{(\mathrm{anc})} \bigg].
\end{aligned}
\end{equation}
After application of the postselection operator $M_0$, in $\circled{1}$ we have
\begin{equation}
\begin{aligned}
    \circled{1} \xrightarrow{M_0} & \ket{\bm{0}}\bra{\bm{0}}P_i^{(\mathrm{sys})}\ket{\bm{0}}\bra{\bm{0}}P_i^{(\mathrm{sys})}\ket{\bm{0}}\bra{\bm{0}} \otimes P_i^{(\mathrm{anc})}\ket{+}\bra{+}P_i^{(\mathrm{anc})} \\
    ={} & \delta^{Z}_i \ket{\bm{0}}\bra{\bm{0}} \otimes P_i^{(\mathrm{anc})}\ket{+}\bra{+}P_i^{(\mathrm{anc})}
\end{aligned}
\end{equation}
where $\delta^{Z}_i$ is zero if $P_i^{(\mathrm{sys})}$ contains any off diagonal Pauli errors $X,Y$ while it is one if $P_i^{(\mathrm{sys})}$ is of $Z$-type only.

Moving onto $\circled{2}$, we find
\begin{equation}
\begin{aligned}
    \circled{2} \xrightarrow{M_0} & \ket{\bm{0}}\bra{\bm{0}}P_i^{(\mathrm{sys})} U^\dag V U \ket{\bm{0}}\bra{\bm{0}} U^\dag V U P_i^{(\mathrm{sys})}\ket{\bm{0}}\bra{\bm{0}} \otimes P_i^{(\mathrm{anc})}\ket{-}\bra{-}P_i^{(\mathrm{anc})} \\
    ={} & |\Gamma_i|^2 \ket{\bm{0}}\bra{\bm{0}} \otimes P_i^{(\mathrm{anc})}\ket{-}\bra{-}P_i^{(\mathrm{anc})}
\end{aligned}
\end{equation}
where we have defined $\Gamma_i = \bra{\bm{0}} U^\dag V U P_i^{(\mathrm{sys})} \ket{\bm{0}}$, which is not quite the desired expectation value $\braket{V}$ due to the erroneous application of a Pauli error $P_i^{(\mathrm{sys})} \ket{\bm{0}}$ resulting in the ensuing overlap calculation. We do note however that $\Gamma_i = \braket{V}$ whenever $P_i^{(\mathrm{sys})}$ is of $Z$-type error since $P_i^{(\mathrm{sys})} \ket{\bm{0}} = \ket{\bm{0}}$ in this case, which will be useful in the following evaluation of terms $\circled{3}$ and $\circled{4}$ as we may write $\delta_i^Z \Gamma_i = \delta_i^Z \overline{\Gamma_i} = \delta_i^Z \braket{V}$. Explicitly,
\begin{equation}
\begin{aligned}
    \circled{3} \xrightarrow{M_0} & \ket{\bm{0}} \underbrace{\bra{\bm{0}}P_i^{(\mathrm{sys})} \ket{\bm{0}}}_{\delta_i^Z} \underbrace{\bra{\bm{0}} U^\dag V U P_i^{(\mathrm{sys})}\ket{\bm{0}}}_{\Gamma_i}\bra{\bm{0}} \otimes P_i^{(\mathrm{anc})}\ket{+}\bra{-}P_i^{(\mathrm{anc})} \\
    ={} & \delta_i^Z\Gamma_i \ket{\bm{0}}\bra{\bm{0}} \otimes P_i^{(\mathrm{anc})}\ket{+}\bra{-}P_i^{(\mathrm{anc})} \\
    ={} & \delta_i^Z\braket{V} \ket{\bm{0}}\bra{\bm{0}} \otimes P_i^{(\mathrm{anc})}\ket{+}\bra{-}P_i^{(\mathrm{anc})}
\end{aligned}
\end{equation}
and
\begin{equation}
\begin{aligned}
    \circled{4} \xrightarrow{M_0} & \ket{\bm{0}} \underbrace{\bra{\bm{0}}P_i^{(\mathrm{sys})} \ket{\bm{0}}}_{\delta_i^Z} \underbrace{\bra{\bm{0}} P_i^{(\mathrm{sys})} U^\dag V U \ket{\bm{0}}}_{\overline{\Gamma_i}}\bra{\bm{0}} \otimes P_i^{(\mathrm{anc})}\ket{-}\bra{+}P_i^{(\mathrm{anc})} \\
    ={} & \delta_i^Z\overline{\Gamma_i} \ket{\bm{0}}\bra{\bm{0}} \otimes P_i^{(\mathrm{anc})}\ket{-}\bra{+}P_i^{(\mathrm{anc})} \\
    ={} & \delta_i^Z\braket{V} \ket{\bm{0}}\bra{\bm{0}} \otimes P_i^{(\mathrm{anc})}\ket{-}\bra{+}P_i^{(\mathrm{anc})}.
\end{aligned}
\end{equation}
Finally, conjugating Equation \eqref{full_phi_rho} with the projection operator $M_0$ yields
\begin{equation}\label{projected_phi_rho}
    M_0\Phi(\rho)M_0 = \ket{\bm{0}}\bra{\bm{0}} \otimes \sum_i \lambda_i P_i^{(\mathrm{anc})} \bigg[ \delta^{Z}_i \ket{+}\bra{+} + |\Gamma_i|^2 \ket{-}\bra{-} + \delta_i^Z \braket{V} \Big( \ket{+}\bra{-} + \ket{-}\bra{+} \Big) \bigg] P_i^{(\mathrm{anc})}.
\end{equation}
and therefore we may read off the final state $\rho^{\mathrm{anc}}$ of the ancilla qubit following the postselection procedure. As a more convenient form, since $P_i^{(\mathrm{anc})} \in \{I,X,Y,Z\}$ we may instead fix some $Q = I,X,Y,Z$ and sum internally over the $2^{2N}$ Pauli operators $P_i \in \{I,X,Y,Z\}^{\otimes (N+1)}$ such that the corresponding $P_i^{(\mathrm{anc})} = Q$. This yields the alternative form for the ancilla density matrix
\begin{equation}\label{collect_anc_rates}
\begin{aligned}
    \rho^{\mathrm{anc}}
    ={} & \sum_{Q \in \{I,X,Y,Z\}} Q \bigg( \sum_{i: P_i^{(\mathrm{anc})}=Q} \lambda_i \rho_i \bigg) Q \\
    ={} & \sum_{i=1}^{\frac{d^2}{4}} \big[ \lambda_i^{I}\rho_i + \lambda_i^{X} X\rho_iX + \lambda_i^{Y} Y\rho_iY + \lambda_i^{Z} Z\rho_iZ \big]
\end{aligned}
\end{equation}
where
\begin{equation}
    \rho_i = \frac{1}{2p_{0|i}} \bigg( \delta_i^{Z} \ket{+}\bra{+} + |\Gamma_i|^2 \ket{-}\bra{-} + \delta_i^{Z} \braket{V} \big[\ket{+}\bra{-} + \ket{-}\bra{+}\big]\bigg).
\end{equation}
Furthermore, by asserting $\Tr{(\rho^{\mathrm{anc}})}=1$ we may infer the postselection probabilities $p_{0|i} = \frac{\delta_i^Z+|\Gamma_i|^2}{2}$ conditional on a particular Pauli error $P_i$ occurring.

\section{Multi-Ancilla Echo Verification}\label{sec:multi_ancilla_EV}

\begin{figure}[b!]
\begin{subfigure}{\linewidth} 
    \centering
    \resizebox{.65\linewidth}{!}{
    \begin{quantikz}[wire types={q,n,q,q,q,q,n,q}, row sep=0.15cm, column sep=0.4cm]
    & \gate{H} & & & & \ctrl{4} & \gate{H}      &  & \meter{\rho^{\mathrm{anc}}} \\
    & \vdots & \vdots & \vdots & \ddots & & \vdots      &  & \vdots \\
    & \gate{H} & & \ctrl{2} & & & \gate{H}      &  & \meter{} \\
    & \gate{H} & \ctrl{1} & & & & \gate{H}      &  & \meter{} \\
    & \gate[4]{U} & \gate[4]{P_1} & \gate[4]{P_2} & & \gate[4]{P_M} & \gate[4]{U^\dag} & \meter{} & \phase{\hspace{2mm}\ket{0}} \setwiretype{c} \wire[u][1]{c} \wire[d][1]{c} \\
    & & & & & & & \meter{} & \phase{\hspace{2mm}\ket{0}} \setwiretype{c} \wire[u][1]{c} \wire[d][1]{c} \\
    \vdots &  & & & \ddots &   & &\vdots & \vdots  \\
    & & & & & & & \meter{} & \phase{\hspace{2mm}\ket{0}} \setwiretype{c} \wire[u][1]{c}
    \end{quantikz}
    }
    \caption{This circuit construction is a simple extension of standard single-ancilla EV, but results in a classical postprocessing overhead that is exponential in the number of ancilla $M$.}
    \label{fig:multi_ancilla_1}
\end{subfigure}
\begin{subfigure}{\linewidth} 
    \centering
    \resizebox{.65\linewidth}{!}{
    \begin{quantikz}[wire types={q,n,q,q,q,q,n,q}, row sep=0.15cm, column sep=0.4cm]
    & \gate{H} & \octrl{1} & \octrl{1} &        & \ctrl{1} & \gate{H}      &  & \meter{\rho^{\mathrm{anc}}} \\
    & \vdots   & \vdots    & \vdots    & \ddots & \vdots   & \vdots      &  & \vdots \\
    & \gate{H} & \octrl{1} & \ctrl{1}  &        & \octrl{1} & \gate{H}      &  & \meter{} \\
    & \gate{H} & \ctrl{1}  & \octrl{1} &        & \octrl{1} & \gate{H}      &  & \meter{} \\
    & \gate[4]{U} & \gate[4]{P_1} & \gate[4]{P_2} & & \gate[4]{P_M} & \gate[4]{U^\dag} & \meter{} & \phase{\hspace{2mm}\ket{0}} \setwiretype{c} \wire[u][1]{c} \wire[d][1]{c} \\
    & & & & & & & \meter{} & \phase{\hspace{2mm}\ket{0}} \setwiretype{c} \wire[u][1]{c} \wire[d][1]{c} \\
    \vdots &  & & & \ddots &   & &\vdots & \vdots  \\
    & & & & & & & \meter{} & \phase{\hspace{2mm}\ket{0}} \setwiretype{c} \wire[u][1]{c}
    \end{quantikz}
    }
    \caption{This multicontrol variant avoids the adverse classical postprocessing overhead encountered for Figure \ref{fig:multi_ancilla_1}, but at the expense of greater complexity in the circuit construction.}
    \label{fig:multi_ancilla_2}
\end{subfigure}
    \caption{Multi-ancilla Echo Verification circuits for the simultaneous estimation of multiple Pauli observables.}
    \label{fig:multi_ancilla}
\end{figure}

In this section we discuss the overhead of multi-ancilla Echo Verification, where each additional ancilla qubit allows a further Pauli observable to be estimated concurrently, and derive two approaches to doing so. Suppose we have a collection of Pauli operators $\{P_m\}_{m=1}^M$ whose expectation values we would like to evaluate on a state $\ket{\psi} = U\ket{\bm{0}}$. The state of the ancilla register produced by the circuit in Figure \ref{fig:multi_ancilla_1} has the form
\begin{equation}
    \ket{\varphi} = \frac{1}{\sqrt{p_{\bm{0}}}} \sum_{\bm{b} \in \{0,1\}^M} \braket{P_{\bm{b}}} \ket{\bm{b}}_{X}
\end{equation}
where $P_{\bm{b}} = \prod_{m=1:b_m=1}^{M} P_m$ is the product of Pauli operators $P_m$ such that the corresponding entry in $\bm{b}$ is equal to one, i.e. $b_m=1$; the subscript $\ket{\cdot}_{X}$ indicates $X$ is the chosen computational basis. Defining by $X_{\bm{a}}$ the Pauli whose $m$-th qubit position is $Z$ when $a_m=1$ and $I$ when $a_m=0$, we obtain
\begin{equation}
    \bra{\varphi} X_{\bm{a}} \ket{\varphi} = \frac{1}{p_{\bm{0}}} \sum_{\bm{b} \in \{0,1\}^M} (-1)^{\bm{a} \cdot \bm{b}} \braket{P_{\bm{b}}}^2.
\end{equation}
The problem is that, in order to isolate a single Pauli term $P_m$ in the above expression, one must sum over all $X_{\bm{a}}$ such that $a_m=1$, of which there are $2^{M-1}$; we note this does not incur any additional coherent resource and is purely classical overhead. Specifically,
\begin{equation}
    \frac{1-\braket{P_m}^2}{p_{\bm{0}}} = \frac{1}{2^{M-1}} \sum_{\bm{a} \in \{0,1\}^M: a_m=1} \bra{\varphi} X_{\bm{a}} \ket{\varphi}.
\end{equation}
A similar expression allows us to evaluate the $Z$ expectation values and consequently the EV estimator; the overall scaling for this approach is $\mathcal{O}(2^{2M-1})$ where $M$ is the number of ancilla.

An alternative approach, which avoids this adverse classical postprocessing overhead, is to use multicontrols on the ancilla register as in Figure \ref{fig:multi_ancilla_2}. Now, it is possible to prepare ancilla states of the form $\ket{\varphi}~=~\frac{1}{\sqrt{p_{\bm{0}}}}~\Big(\ket{0}~+~\sum_{m=1}^M~\braket{P_m} \ket{m}\Big)$ and extract the relevant expectation values with ease. However, the added gate-cost is considerable and not appropriate for near-term quantum devices; in short, the two approaches discussed in this Section offer a trade-off between easy circuit construction with exponential classical post-processing, and low-cost processing with more complex circuit structure.

\end{document}